\documentclass[10pt,a4paper]{article}
\pdfoutput=1
\usepackage[english]{babel}
\usepackage{longtable}
\usepackage{amsmath,amsfonts,amssymb,bm}\interdisplaylinepenalty=2500

\newcommand{\unit}[1]{\ensuremath{\mathrm{#1}}}

\newcommand{\ohm}{\ensuremath{\mathrm{\Omega}}}
\newcommand{\req}[1]{(\ref{#1})}
\usepackage{graphicx,color,rotating,epstopdf}
\def\PrintGraphicFileName{1}			% 0/1, user defined
\newcommand{\namedgraphics}[3]{%     % #1=scale #2=file #3=\textwidth or \columnwidth
	\parbox{#3}{%
	\ifnum\PrintGraphicFileName>0\rotatebox{90}{\smash{\ttfamily\scriptsize\raisebox{0.8em}{#2}}}\fi%
	\hspace*{\fill}\includegraphics[scale=#1]{#2}\hspace*{\fill}}}
 \newcommand{\namedcombographics}[2]{%   % Combined Latex eps/pdf
	\parbox{\textwidth}{%
	 \ifnum\PrintGraphicFileName>0\rotatebox{90}{\smash{\ttfamily\scriptsize\raisebox{0.8em}{#2}}}\fi%
	\hspace*{\fill}\scalebox{#1}{%  %              % scale
	\ifnum\pdfoutput=0\input{\LatexGraphicsPath#2.pstex_t}
	\else\input{\LatexGraphicsPath#2.pdftex_t}\fi}\hspace*{\fill}}}
\newcommand{\namedlatexgraphics}[2]{%     % #1=scale #2=file
	\parbox{\textwidth}{%
	 \ifnum\PrintGraphicFileName>0\rotatebox{90}{\smash{\ttfamily\scriptsize\raisebox{0.8em}{#2}}}\fi%
	\hspace*{\fill}\scalebox{#1}{%  %              % scale
	\input{\LatexGraphicsPath#2.latex}}\hspace*{\fill}}}
\title{Applications of the optical fiber to the generation and to the measurement of low-phase-noise microwave signals}
\author{K. Volyanskiy\thanks{K. V. is also with the St.\ Petersburg State University of Aerospace Instrumentation, Russia.},
J. Cussey\thanks{Now with Smart Quantum, Lannion \& Besan\c{c}on, France.},
H. Tavernier,\\
P. Salzenstein,
G. Sauvage\thanks{Aeroflex, Elancourt (Paris), France. Web site http://www.aeroflex.com},
L. Larger,
E. Rubiola\thanks{corresponding author, e-mail rubiola@femto-st.fr, home page http://rubiola.org}\\
\small web page \texttt{http://rubiola.org}
\\[4em]\includegraphics[width=0.35\textwidth]{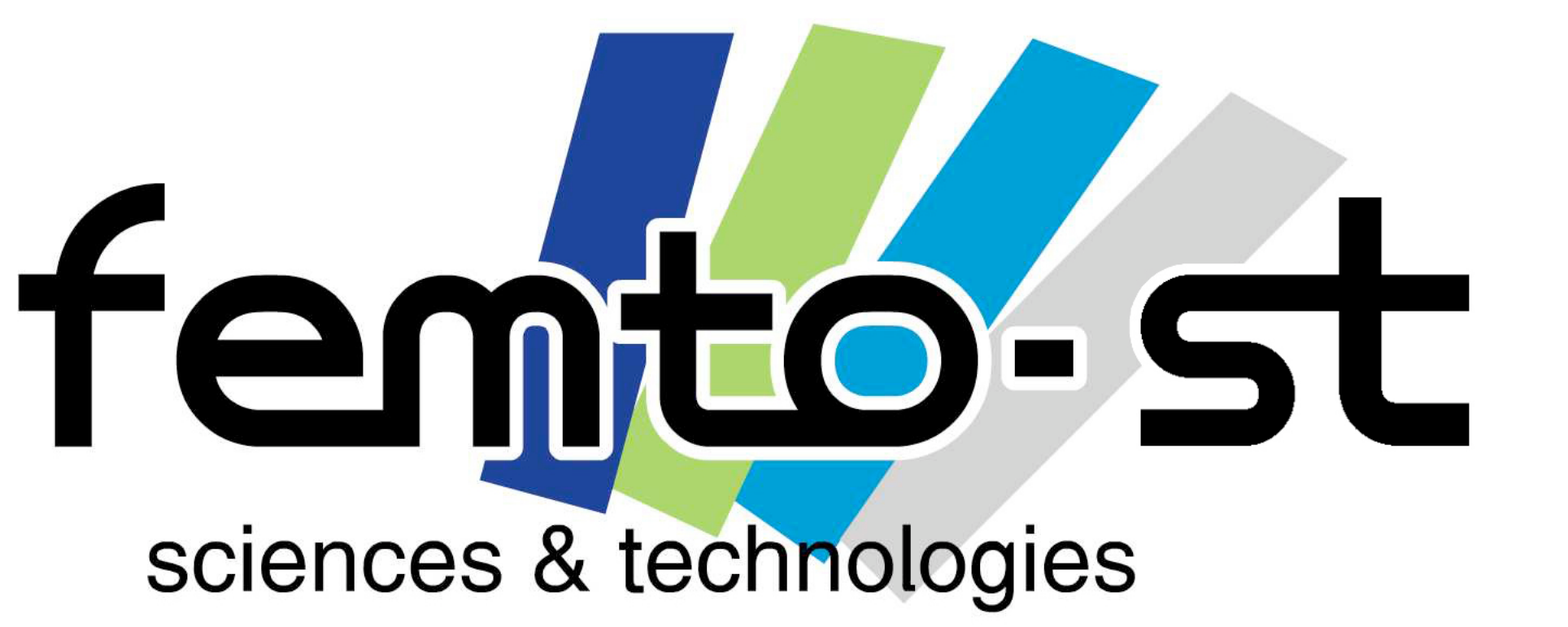}\\[0.5em]
\small FEMTO-ST Institute\\[-0.5ex]
\small CNRS and Universit\'e de Franche Comt\'e, 
\small Besan\c{c}on, France\\[1.5em]}
\date{\small\today}
\def\myheaders{E. Rubiola \& al.,\quad Optical fiber \& microwave phase noise\hfill\today\quad}
\begin{document}
\maketitle

\begin{abstract}
The optical fiber used as a microwave delay line exhibits high stability and low noise and makes accessible a long delay ($\gtrsim100$ $\mu$s) in a wide bandwidth ($\approx40$ GHz, limited by the optronic components).  Hence, it finds applications as the frequency reference in microwave oscillators and as the reference discriminator for the measurement of phase noise.  The fiber is suitable to measure the oscillator stability with a sensitivity of parts in $10^{-12}$.  
Enhanced sensitivity is obtained with two independent delay lines, after correlating and averaging. 
Short-term stability of parts in $10^{-12}$ is achieved inserting the delay line in an oscillator.  The frequency can be set in steps multiple of the inverse delay, which is in the 10--100 kHz region. 

This article 
%stands on two references [E. Rubiola \& al.\ JOSAB 22(5) 987, (2005), X. S. Yao \& al.\ JOSAB 13(8) 1725 (1996)], adding 
adds to the available references a considerable amount of engineering and practical knowledge, the understanding of $1/f$ noise, calibration, the analysis of the cross-spectrum technique to reduce the instrument background, the phase-noise model of the oscillator, and the experimental test of the oscillator model.
\end{abstract}

\def\SymbolTable{%
\begin{center}
\addcontentsline{toc}{section}{Most used symbols}
\begin{longtable}[t]{ll}
\multicolumn{2}{l}{\large\bfseries Most used symbols}\\\hline
\bfseries symbol	& \bfseries meaning\\\hline
A			& amplifier voltage gain (constant)\\
%$a(t)\leftrightarrow A(jf)$ & single-channel random signal\\
%$b(t)\leftrightarrow B(jf)$ & single-channel random signal\\
$b_i$	& coefficients of the power-law approximation of $S_\varphi(f)$\\
$\mathrm{B}(jf)$ & phase response of the oscillator's feedback path\\
%$c(t)\leftrightarrow C(jf)$	& random signal common to the two channels\\
$f$		& Fourier (or near-dc) frequency, Hz\\
$F$		& amplifier noise figure\\
$h$		& Planck constant, $h=6.626{\times}10^{-19}$ J\:s\\
$h_i$	& coefficients of the power-law approximation of $S_y(f)$\\
$H(jf)$	& transfer function (optical delay line)\\ 
$\mathrm{H}(jf)$ & phase-noise transfer function (oscillator)\\
$i(t)$		& current, as a function of time \\
$j$		& imaginary unit, $j^2=-1$ \\
$J_n(z)$	& Bessel function of order $n$\\
$k$		& Boltzmann constant, $k=1.381{\times}10^{-23}$ J/K\\
$k_\varphi$ & phase-to-voltage gain (of the phase detector)\\
$m$		& integer number (in averages)\\
$m_i$, $m_\varphi$	& intensity-modulation index, phase-modulation index\\
$N$		& noise power spectral density, W/Hz\\
$P_0$, $P_\lambda$ & microwave power, optical power\\
$q$		& electron charge, $q=1.602{\times}10^{-19}$ C\\
$Q$		& resonator quality factor\\
$R_0$	& load resistance ($R_0=50$ \ohm)\\
$S_x(f)$	& power spectral density (PSD) of $x(t)$\\ 
$t$		& time\\
$T_0$	& reference temperature, $T_0=290$ K\\
$v(t)$	& voltage, as a function of time\\
$x(t)\leftrightarrow X(jf)$& a variable (in correlation measurements)\\
$y(t)$	& fractional frequency fluctuation\\
$y(t)\leftrightarrow Y(jf)$& a variable (in correlation measurements)\\
$V$, $V_0$ & dc or peak voltage\\
$\alpha(t)$	& normalized-amplitude noise\\
$\beta(jf)$	& transfer function of the feedback path\\
$\Delta f$	& peak frequency deviation (in frequency modulation)\\
%$\eta$	& photodetector quantum efficiency\\
$\lambda$	& optical wavelength\\
$\nu_0$	& carrier frequency, Hz\\
$\rho$	& photodetector responsivity, A/W\\
$\sigma_y(\tau)$	& Allan deviation, 
		square root of the Allan variance $\sigma^2_y(\tau)$\\
$\tau$	& measurement time, in $\sigma_y(\tau)$\\
$\tau$	& delay of a delay line (instrument)\\
$\tau_d$, $\tau_f$ & in oscillator, delay of the line and of the selector filter\\
$\varphi(t)\leftrightarrow\Phi(jf)$ & phase noise or modulation\\
$\psi(t)\leftrightarrow\Psi(jf)$ & amplifier phase noise\\[1em]
[uppercase]	& Fourier transform, as in $x(t)\leftrightarrow X(jf)$\\
$\leftrightarrow$& transform inverse-transform pair, as in 
	$x(t)\leftrightarrow X(jf)$\\
$<~>_m$	& mean of $m$ values\\
$\overline{x}$	& time average of $x(t)$
\end{longtable}
\end{center}}

\clearpage\tableofcontents\clearpage\SymbolTable\clearpage

\section{Introduction}
Science and technology require reference microwave sources with ever increasing stability and spectral purity, and of course suitable measurement systems.  The frequency synthesis from the HF/VHF region is no longer satisfactory because of the insufficient spectral purity inherent in the multiplication, and because of more trivial limitations like mass, volume and complexity.  
The domain of RF/microwave photonics is growing fast \cite{chang:rf-photonics}.
The generation of microwaves from optics, or more generally `with optics,' is providing new solutions.  Examples are mode-locked lasers \cite{delfyett03jqe,ippen03ol}, opto-electronic oscillators photonic \cite{yao96josab,yao00jlt} (OEO), optical micro-cavities \cite{vahala03nature}, optical parametric oscillators \cite{kippenberg04prl,savchenkov04prl} (OPO), and frequency combs \cite{ye03rmp}.

The optical fiber used as a delay line enables the generation \cite{yao96josab,yao00jlt} and the measurement \cite{rubiola05josab-delay-line} of stable and highly spectrally pure microwave signals.  The optical fiber is a good choice for the following reasons.
\begin{enumerate} 
\item A long delay can be achieved, of 100 $\mu$s and more, thanks to the low loss (0.2 dB/km at 1.55 $\mu$m and 0.35 dB/km at 1.31 $\mu$m).
\item The frequency range is wide, at least of 40 GHz, still limited by the opto-electronic components
\item The background noise is low, close to the limit imposed by the shot noise and by the thermal noise at the detector output.
\item The thermal sensitivity of the delay ($6.85{\times}10^{-6}$/K) is a factor of 10 lower than the sapphire dielectric cavity at room temperature.  This resonator is considered best ultra-stable microwave reference.
\item In oscillators and phase-noise measurements, the microwave frequency is the inverse of the delay.  This means that the oscillator or the instrument can be tuned in step of $10^{-5}$--$10^{-6}$ of the carrier frequency without degrading stability and spectral purity with frequency synthesis.  Finer tuning is possible at a minimum cost in terms of stability and spectral purity. 
\end{enumerate}
We describe a series of experiments related to the generation and to the measurement of low-phase-noise microwave signals using a  delay implemented with an intensity-modulated beam propagating through an optical fiber.  A 10 GHz oscillator prototype exhibits a frequency flicker of $3.7{\times}10^{-12}$ (Allan deviation) and a phase noise lower than $-140$ \unit{dBrad^2/Hz} at 10 kHz off the carrier.  The same values are achieved as the sensitivity in phase noise measurements in real time.  This sensitivity is sufficient to measure the frequency flicker of a sapphire oscillator, which is considered the reference in the field of low-noise microwave oscillators \cite{giordano05epjap-oscillators}.
In phase noise measurements, enhanced sensitivity is obtained with the cross-spectrum method, which takes correlation and averaging on two independent delay lines. 

This article stands on Ref.~\cite{yao96josab} for the oscillator, and on  Ref.~\cite{rubiola05josab-delay-line,salik04fcs-xhomodyne} for the measurements.  We aim at providing practical knowledge, adding engineering, accurate calibration, the analysis of the cross-spectrum technique, the phase-noise model of the oscillator, and experimental tests.

\section{Basic concepts}
%===================================

\subsection{Phase noise}\label{ssec:ddl-phase-noise}
%------------------------------------------------------------
Phase noise is a well established subject, clearly explained in classical references, among which we prefer \cite{rutman78pieee,kimball:precise-frequency,ccir90rep580-3,ieee99std1139} and \cite[vol.\,1, chap.\,2]{vanier:frequency-standards}.
The quasi-perfect sinusoidal signal of frequency $\nu_0$, of random amplitude fluctuation $\alpha(t)$, and of random phase fluctuation $\varphi(t)$ is
\begin{align}
v(t)=\left[1+\alpha(t)\right]\cos\left[2\pi\nu_0t+\varphi(t)\right]~.
\label{eqn:ddl-clock-signal}
\end{align}
We may need that $|\alpha(t)|\ll1$ and $|\varphi(t)|\ll1$ during the measurement.
The phase noise is generally measured as the average PSD (power spectral density)
\begin{align}
S_{\varphi}(f) = \left<|\Phi(jf)\}|^2\right>_m
&&\text{(avg, $m$ spectra)},
\end{align}
The uppercase denotes the Fourier transform, so $\varphi(t)\leftrightarrow\Phi(jf)$ form a transform inverse-transform pair.
In experimental science, the single-sided PSD is preferred to the two-sided PSD because the negative frequencies are redundant.
It has been found that the power-law model describes accurately the phase noise of oscillator and components
\begin{align}
&S_{\varphi}(f) = \sum_{n=-4}^{0}b_nf^n \qquad\qquad\text{(power law)}
\label{eqn:ddl-power-law-sphi}\\
&\text{\begin{tabular}{cl}
coefficient & noise type\\\hline
$b_{-4}$ & frequency random walk\\
$b_{-3}$ & flicker of frequency\\
$b_{-2}$ & white frequency noise, or phase random walk\\
$b_{-1}$ & flicker of phase\\
$b_{0}$ & white phase noise
\end{tabular}}\nonumber
\end{align}
The power law relies on the fact that white ($f^0$) and flicker ($1/f$) noises exist per-se, and that a phase integration is present in oscillators, which multiplies the spectrum $\times1/f^2$.  If needed, the model can be extended to $n<-4$.
Of course, the power law can also be used to describe the spectrum the fractional frequency fluctuation $y(t)=\dot{\varphi}(t)/2\pi\nu_0$  
\begin{align}
S_{y}(f)=\frac{f^2}{\nu_0^2}\:S_{\varphi}(f) = \sum_{n=-2}^{2}h_nf^n~.
\label{eqn:ddl-sy}
\end{align}

Another tool often used is the two-sample (Allan) variance $\sigma^2_y(\tau)$ as a function of the measurement time $\tau$.  Notice that the symbol $\tau$ is commonly used for the measurement time and for delay of the line (Sec.~\ref{ssec:ddl-delay-line-theory}).  We will added a subscript when needed.  For the most useful frequency-noise processes, the relation between $\sigma^2_y(\tau)$ and $S_y(f)$ is
\begin{align}
\sigma^2_y(\tau)=\begin{cases} 
	\displaystyle
	\frac{h_0}{2\tau}	& \text{white frequency noise}\\[2ex]
	h_{-1}\,2\ln(2)		& \text{flicker of frequency}\\[1ex]
	\displaystyle
	h_{-2}\,\frac{(2\pi)^2}{6}\,\tau	& \text{frequency random walk}
	\end{cases}
\label{eqn:ddl-sy-sigmay}
\end{align}

\subsection{Cross-spectrum method}\label{ssec:ddl-xspectrum}
%------------------------------------------------------------
Inevitably, the measured noise is the sum of  the device-under-test  (DUT) noise and of the the instrument background.  Improved sensitivity is obtained with a correlation instrument, in which two separate channels measure simultaneously the same DUT\@.
Let $a(t)\leftrightarrow A(jf)$ and $b(t)\leftrightarrow B(jf)$ the background of the two instruments, and $c(t)\leftrightarrow C(jf)$ the DUT noise or any common noise.  The two instrument outputs are 
\begin{align}
x(t)&=c(t)+a(t)\\
y(t)&=c(t)+b(t)~,
\end{align}
where $a(t)$, $b(t)$ and $c(t)$ are statistically independent because we have put all the common noise in $c(t)$.  The cross-spectrum averaged on $m$ measures is
\begin{align}
S_{yx}(f) 
&=\left<YX^\ast\right>_m \nonumber\\
%&=\left<\{[C+A]\times[C+B]^\ast\right>_m \nonumber\\
&=\left<CC^\ast\right>_m + \left<CB^\ast\right>_m 
	+ \left<AC^\ast\right>_m + \left<AB^\ast\right>_m \nonumber\\
&= S_{c}(f) + O(\sqrt{1/m})~,
\label{eqn:ddl-Syx-avg} 
\end{align}
where $O(\:)$ means `order of.'  Owing to statistical independence of $a(t)$, $b(t)$ and $c(t)$, also $A(jf)$, $B(jf)$ and $C(jf)$ are statistically independent.  Hence, the cross terms decrease as $\smash{\sqrt{1/m}}$.  This enables to interpret $S_{yx}(f)$ as follows.
\begin{description}
\item[Statistical limit.] With no DUT noise and with two fully independent channels, it holds that $c(t)=0$.  After \req{eqn:ddl-Syx-avg}, the statistical limit of the measurement is
\begin{align}
S_{yx}(f) \approx\sqrt{\frac1m\,S_a(f)\,S_b(f)}
\qquad\text{(statistical limit)}.
\end{align}
Accordingly, a 5 dB improvement on the single-channel noise costs a factor of 10 in averaging, thus in measurement time.

\item[Correlated hardware background.] Still at zero DUT noise, we break the hypothesis of statistical independence of the two channels.  We interpret $c(t)$ as the correlated noise of the instrument, due to environment, to the crosstalk between the two channel, etc.  This is the hardware limit of the instrument sensitivity
\begin{align}
S_{yx}(f) \simeq [S_{c}(f)]_\text{xtalk,\,etc.}
\qquad\text{(hardware limit)}.
\label{eqn:ddl-correl-hw-limit}
\end{align}

\item[Regular DUT measurement.] Now we introduce the DUT noise. Under the assumptions that\vspace{-1.5ex} 
\begin{enumerate}\addtolength{\itemsep}{-0.8ex}
\item $m$ is large enough for the statistical limit to be negligible, and 
\item the single-channel background is negligible as compared to the DUT noise,
\end{enumerate}\vspace{-1.5ex} 
the cross spectrum gives the \emph{DUT noise}
\begin{align}
S_{yx}(f) \simeq [S_{c}(f)]_\text{DUT}
\qquad\text{(DUT measurement).}
\label{eqn:ddl-correl-dut-meas}
\end{align}
This is the regular use of the instrument.

\end{description}

\subsection{Delay line theory}\label{ssec:ddl-delay-line-theory}
%-------------------------------------------------------------
\begin{figure}[t]
\centering\namedgraphics{0.65}{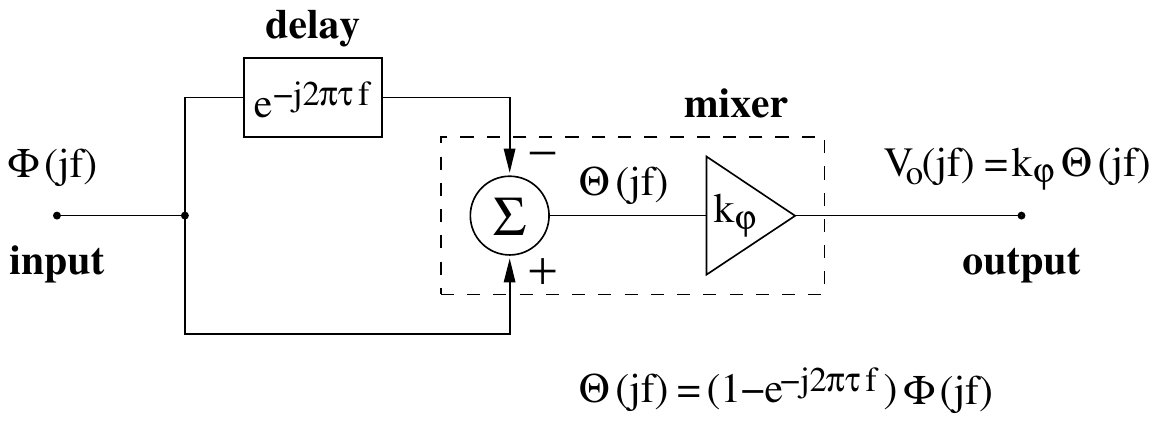}{\columnwidth}
\caption{Basic delay-line phase noise measurement.}
\label{fig:ddl-delay-basic}
\end{figure}
Delaying the signal $v(t)$ by $\tau$, all time-varying parameters of $v(t)$ are also delayed by $\tau$, thus the phase fluctuation $\varphi(t)$ turns into $\varphi(t-\tau)$.
By virtue of the time-shift theorem, the Fourier transform of $\varphi(t-\tau)$ is $\smash{e^{-j2\pi\tau f}}\Phi(jf)$.
This enables the measurement of the oscillator phase noise $\varphi(t)$ by observing the difference $\theta(t)=\varphi(t)-\varphi(t-\tau)$.  
Referring to Fig.~\ref{fig:ddl-delay-basic}, the output signal is 
$V_o(jf)=k_\varphi\Theta(jf)=k_\varphi H(jf)\Phi(jf)$,
where $k_\varphi$ is the mixer phase-to-voltage gain, and $H(jf)=1-e^{-j2\pi\tau f}$ is the system transfer function.  Consequently
\begin{gather}
S_{v}(f)=k_\varphi^2 |H(jf)|^2S_{\varphi}(f)
\label{eqn:ddl-sphi-o}\\
|H(jf)|^2=4\sin^2(\pi f\tau)~.
\label{eqn:ddl-hphi2}
\end{gather}
The oscillator noise $S_{\varphi}(f)$ is inferred by inverting \req{eqn:ddl-sphi-o}.  
In practice, it is important to keep $S_v(f)$ accessible because it reveals most of the experimental mistakes connected with the instrument background.
The function $|H(jf)|^2$ has a series of zeros at $f=n/\tau$, integer $n$, in the vicinity of which the experimental results are not useful.  In practice, the first zero sets the maximum measurement bandwidth to $0.95/\tau$, as discussed in \cite{rubiola05josab-delay-line}.  
Nonetheless, the regions between contiguous zeros are useful diagnostics for the oscillator under test, provided the frequency resolution of the FFT analyzer be sufficient.

At low frequency, the instrument background is naturally optimized for the measurement of $1/f^2$ and $1/f^3$ noise because $1/|H(jf)|^2\sim 1/f^2$ for $f\rightarrow0$, and an additional factor $1/f$ comes from the electronics.
This is a fortunate outcome because the noise of the oscillators of major interest is proportional to $1/f^3$ or to $1/f^4$ in this region.
Of course, an appropriate choice of $\tau$ is necessary.

\subsection{Oscillator phase noise}\label{ssec:ddl-oscillator-noise}
%------------------------------------------------------------------------
\begin{figure}[t]
\centering\namedgraphics{0.8}{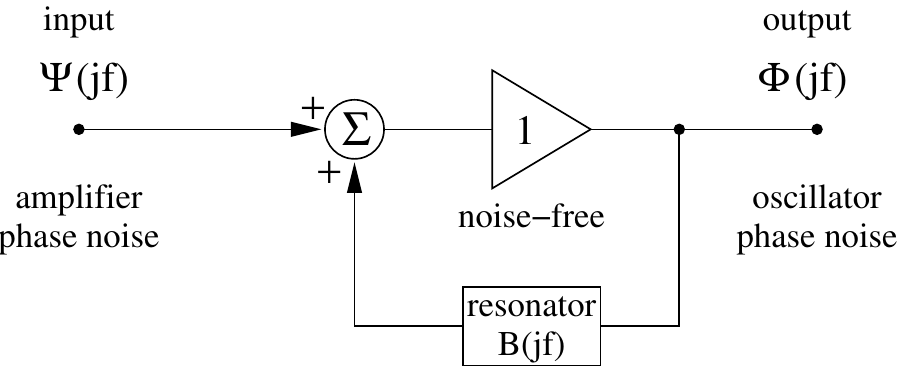}{\columnwidth}
\caption{Oscillator phase-noise model.}
\label{fig:ddl-phase-xfer}
\end{figure}

The oscillator consists of an amplifier of gain $A$ (assumed constant vs.\ frequency) and of a feedback transfer function $\beta(jf)$ in closed loop.  The gain $A$ compensates for the losses, while $\beta(jf)$ selects the oscillation frequency.  This model is general, independent of the nature of $A$ and $\beta(jf)$.
We assume that the Barkhausen condition $|A\beta(jf)|=1$ for stationary oscillation is verified at the carrier frequency $\nu_0$ by saturation in the amplifier or by some other gain-control mechanism.  Under this hypothesis, the phase noise is modeled by the scheme of Fig.~\ref{fig:ddl-phase-xfer}, where all signals are \emph{the phases of the oscillator loop} \cite{leeson-arxiv,leeson-cambridge}.  This model is inherently linear, so it eliminates the mathematical difficulty due to the parametric nature of flicker noise and of the noise originated from the environment fluctuations.
We denote with $\varphi(t)\leftrightarrow\Phi(jf)$ the oscillator phase noise, and with $\psi(t)\leftrightarrow\Psi(jf)$ the amplifier phase noise. The latter is always additive, regardless of the underlying physical mechanism.  More generally, $\psi(t)$ accounts for the phase noise of all the electronic and optical components in the loop. 
The ideal amplifier `repeats' the phase of the input, for it has a gain of one (exact) in the phase-noise model.
The feedback path is described by the transfer function $\mathrm{B}(jf)$ of the phase perturbation.  
In the case of the \emph{delay-line oscillator}, the feedback path is a delay line of delay $\tau_d$ followed by a resonator of relaxation time $\tau_f\ll\tau_d$ that selects the oscillation frequency $\nu_0$ among the multiples of $1/\tau_d$.  Neglecting the difference between natural frequency and oscillation  frequency, $\tau_f$ is related to the quality factor $Q$ by $\tau_f=\frac{Q}{\pi\nu_0}$.  The phase-perturbation response of the feedback path is  
\begin{align}
\mathrm{B}(jf)=\frac{e^{-j2\pi f\tau_d}}{1+j2\pi f\tau_f}~.
\label{eqn:ddl-B}
\end{align}
The oscillator is described by the phase noise transfer function 
\begin{align}
\mathrm{H}(jf) = \frac{\Phi(jf)}{\Psi(jf)}
\qquad\text{[definition of $\mathrm{H}(jf)$]}~.
\end{align}
Using the basic equations of feedback, by inspection on Fig.~\ref{fig:ddl-phase-xfer} we find 
\begin{gather}
\mathrm{H}(jf)=\frac{1}{1-\mathrm{B}(jf)}~,
\intertext{and consequently}
%\mathrm{H}(jf)=\frac{1+j2\pi f\tau_f}{1+j2\pi f\tau_f-e^{-j2\pi f\tau_d}}~.
\left|\mathrm{H}(jf)\right|^2=\frac{1+4\pi^2f^2\tau_f^2}{2-2\cos(2\pi f\tau_d)+4\pi^2f^2\tau_f^2+2\omega\tau_f\sin(2\pi f\tau_d)}~.
\label{eqn:ddl-oscillator}
\end{gather}
The detailed proof of \req {eqn:ddl-B} and \req{eqn:ddl-oscillator} is given in \cite{leeson-arxiv,leeson-cambridge}.  The result is confirmed using the phase diffusion and the formalism of stochastic processes \cite{chembo08jqe-oeo-phase-diffusion}.
\begin{figure}[t]
\centering\namedgraphics{0.68}{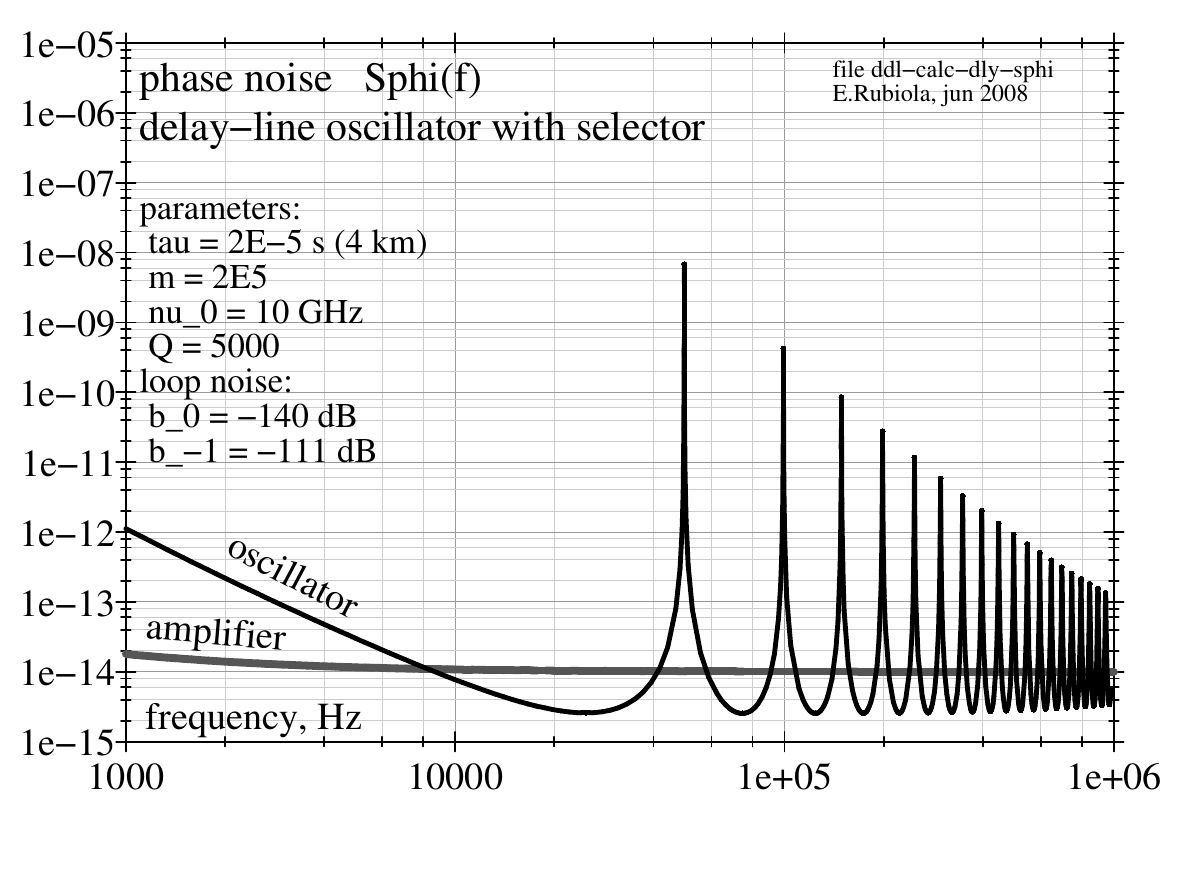}{\columnwidth}
\vspace*{-5ex}
\caption{Calculated OEO phase-noise.}
\label{fig:ddl-calc-dly-sphi}
\end{figure}
Figure \ref{fig:ddl-calc-dly-sphi} shows an example of OEO phase-noise.  The loop noise is $S_\psi(f)=8{\times}10^{-12}/f+10^{-14}$, the same used in Sec.~\ref{ssec:oeo-prototype}.  The periodicity of the delay-line phase produces a series of noise peaks at frequency multiple of $1/\tau_d$.  These peaks have extremely narrow bandwidth, in the Hertz range, but in simulations and experiments they seem significantly wider because of the insufficient resolution.  The peak height follows the law $S_\varphi(f)\sim1/f^4$ ($-40$ dB/decade).  Eliyahu \cite[Sec.\,IV]{eliyahu03fcs} reports about a discrepancy by a factor $\sqrt{f}$ ($-35$ dB/decade), yet giving little details and suggesting that further investigation is necessary.

\section{Optical-fiber microwave delay}
%==================================
\subsection{Design strategy}
%--------------------------------------------------------------
\begin{figure*}[t]
\centering\namedgraphics{0.75}{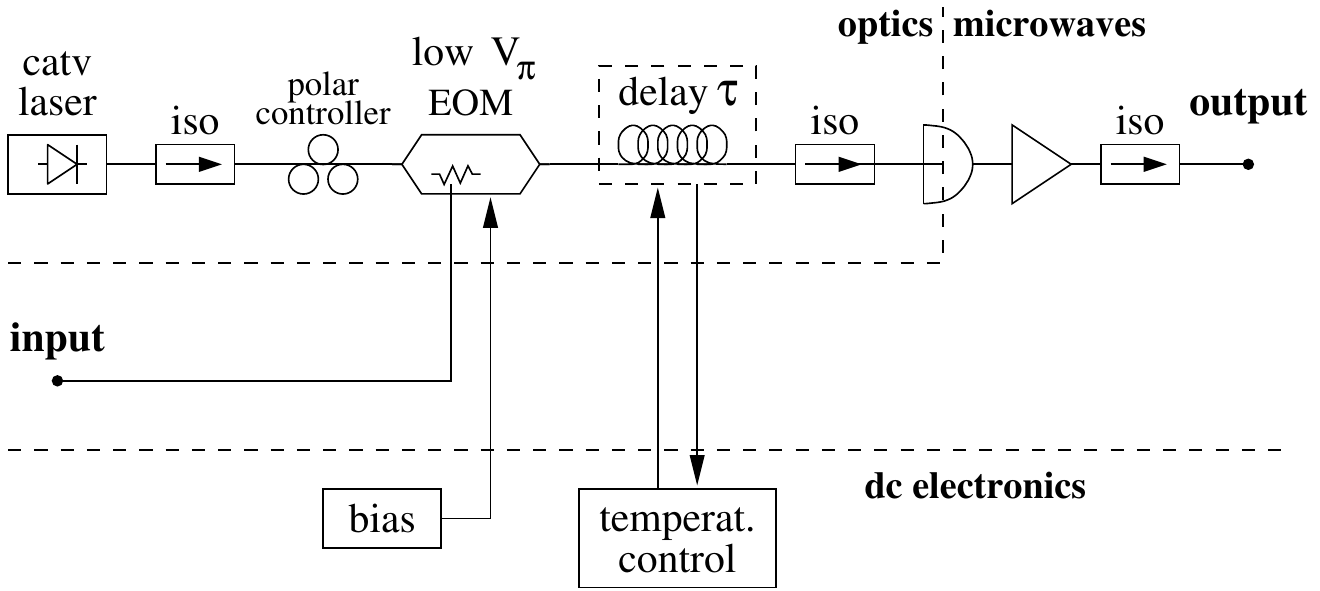}{\textwidth}
\caption{Optical-fiber delay unit.}
\label{fig:ddl-photonic-channel}
\end{figure*}
The optical-fiber delay unit, shown in Fig.~\ref{fig:ddl-photonic-channel}, is designed for low noise and for high stability of the delay, as discussed underneath.  Focusing on the X band, for practical reasons we chose standard 7--12.4 GHz microwave components, mainly intended for radar and telecommunication technology.  

The optical fiber is a Corning SMF-28 wound on a cylinder of 15 cm diameter and 2 cm heigth.  We used 2 km (10 $\mu$s) and 4 km (20 $\mu$s) fibers in most experiments, and sometimes shorter ones.  The spool is enclosed in a 5 mm thick Duralumin cylinder thermally insulated from the environment by a 3 cm plastic foam layer.  The cylinder is stabilized at room temperature within a fraction of a milliKelvin with a PID control built in our laboratory, and set with the Ziegler-Nichols method.  The ultimate period, i.e., the inverse oscillation frequency of the proportional-only control at the oscillation threshold, is of 40 s.  The advantage of the control vs.\ a passive stabilization (large metal mass and insulator) is still questionable.  In the short-term the passive stabilization would certainly be preferable because it does not suffer from the noise inherent in the control.  On the other hand, in phase noise measurements (Fig.~\ref{fig:ddl-correl-instrum}) we need to stabilize the quadrature condition at the mixer inputs during the session, which last up to one day.  Moreover, in dual-channel measurements (Section \ref{sec:dual-channel}) the residual environment fluctuations are fully correlated, while the thermal fluctuations of the control can be rejected because the two controls are independent.

The light source is a semiconductor CATV laser, temperature controlled and powered with a low-noise current source.  This choice is partially motivated by the need of reasonable simplicity and of low RIN (relative intensity noise).  The RIN turns into AM noise of the detected microwave signal, which pollutes the phase noise measurements (Section \ref{ssec:ddl-mixer-noise}).
During the past two years we used both 1.31 $\mu$m and 1.55 $\mu$m lasers.   The fiber attenuation is of 0.2 dB/km at 1.55 $\mu$m, and of  0.35 dB/km at 1.31 $\mu$m. 
The dispersion of the SMF-28 fiber goes to zero at $1.311\pm0.01$ $\mu$m, which virtually eliminates the effect of the laser frequency noise in the vicinity of that wavelength.  For reference, at 1.55 $\mu$m the dispersion is of 17 ps/nm/km.  A laser linewidth of 10 MHz ($5.8{\times}10^{-5}$ nm) at 1.55 $\mu$m produces a delay fluctuation of $2{\times}10^{-15}$ s rms after 2 km optical fiber,  which is equivalent to $1.2{\times}10^{-4}$ rad at the microwave frequency of 10 GHz.
The frequency flicker of one of our lasers, measured in the frequency domain with an asymmetric Mach-Zehnder and converted into Allan deviation $\sigma_y(\tau)$, is of $4{\times}10^{-10}$.  This preliminary result indicates that the laser fluctuation should be of less than 100 kHz (flicker floor), which gives a phase-noise contribution lower than other noises.  In the end, we have a weak preference for the 1.55 $\mu$m lasers, based on the more progressed technology, and after comparing empirically the effect several lasers on the phase noise spectra.
 
The intensity modulator is a Mach-Zehnder electro-optic modulator (EOM) exhibiting low half-wave voltage ($V_\pi\simeq3.9$ V), so that the maximum modulation is achieved with no more than 50 mW ($+17$ dBm) of microwave power.  This choice is important for the stability of the bias point because the \unit{LiNbO_3} is highly sensitive to temperature, thus to power and to thermal gradients.  Other modulation methods have been discarded, the direct modulation of the laser because the laser threshold enhances the microwave phase noise, and the acusto-optic modulator because it is unsuitable to microwaves.

For low noise, the photodetector can only be a InGaAs p-i-n diode operated in strong reverse-bias conditions, thus a photoconductor.  Reverse bias is necessary for high speed, as it reduces the capacitance.  The need for low noise excludes some other detectors, like the avalanche diode.  The traditional photodetectors loaded to a resistor are preferred to the more modern ones with internal transconductance amplifier because of the possibility to choose a low-flicker external amplifier.

\subsection{Output power and white noise}
%---------------------------------------------------------
Using the subscript $\lambda$ for light and the overline for the time-average, the modulated optical power at the output of the EOM is 
\begin{align}
P_\lambda(t)=\overline{P}_\lambda\left[1+m_i\cos(2\pi\nu_0 t)\right]~,
\end{align}
where the intensity-modulation index is 
\begin{align}
m_i=2J_1\left(\frac{\pi V_p}{V_\pi}\right)~,
\label{eqn:ddl-mz-bessel}
\end{align}
$J_1(\:)$ is the 1st order Bessel function, $V_p$ is the microwave peak voltage, and $V_\pi$ is the modulator half-wave voltage.  Equation \req{eqn:ddl-mz-bessel} originates from the sinusoidal nature of the EOM response, with $V_p\cos(2\pi\nu_0t)$ input voltage.  The harmonics at frequency $n\nu_0$, integer $n\ge2$, fall beyond the microwave bandwidth, thus they are discarded.  Though the maximum modulation index is $m_i=1.164$, occurring at $V_p=0.586\,V_\pi$, the practical values are 0.8--1.

The detected photocurrent is $i(t)=\rho P_\lambda(t)$, where $\rho$ is the detector responsivity.  Assuming a quantum efficiency of 0.6, the responsivity is of 0.75 A/W at 1.55 $\mu$m wavelength, and of 0.64 A/W at 1.31 $\mu$m.
The dc component of $i(t)$ is $i_\text{dc}=\rho\overline{P}_\lambda$.  The microwave power at the detector output is
\begin{align}
P_0=\frac{1}{2}m_i^2R_0\rho^2\overline{P}^2_\lambda
\qquad\text{(detector output)}~,
\end{align}
where $R_0=50$ \ohm\ is the load resistance.

The white noise at the input of the amplifier is
\begin{align}
N = FkT_0 + 2qR_0\rho\overline{P}_\lambda
\qquad\text{(white noise)}~,
\label{eqn:ddl-total-noise}
\end{align}
where $F$ is the amplifier noise figure and $kT_0=4{\times}10^{-21}$ J is the thermal energy at room temperature.  The first term of \req{eqn:ddl-total-noise} is the noise of the amplifier, and the second term is the shot noise.
Using $b_{0}=N/P_0$, the white phase noise is
\begin{gather}
b_0=\frac{2}{m_i^2}\left[
	\frac{FkT_0}{R_0} \: \frac{1}{\rho^2\overline{P}_\lambda^2} +
        \frac{2q}{\rho\overline{P}_\lambda}
         \right]
         \qquad\text{(white phase noise)}~.
\end{gather}
Interestingly, the noise floor is proportional to 
$1/\overline{P}_\lambda^2$ at low power, and to
$1/\overline{P}_\lambda$ above the threshold power
\begin{align}
P_{\lambda}=\frac{FkT_0}{R_0} \: \frac{1}{2\rho q}
\qquad\text{(threshold power)}~.
\end{align}
For example, taking $\rho=0.75$ A/W and $F=5$ (SiGe parallel amplifier), we get a threshold of 1.7 mW, at which the noise floor is  $b_0\simeq10^{-15}$ \unit{rad^2/Hz} ($-150$ \unit{dBrad^2/Hz}).

\subsection{Flicker noise}
%---------------------------------------------------------------------
Phase and amplitude flickering result from the near-dc $1/f$ noise up-converted by non-linearity or by a parametric modulation process.  
This is made evident by two simple facts:\vspace{-1.5ex} 
\begin{enumerate}\addtolength{\itemsep}{-0.8ex}
\item The $1/f$ noise is always always present in the dc bias of electronic devices.
\item In the absence of a carrier, the microwave spectrum at the output of a device is white, i.e.\ nearly constant in a wide frequency range.
\end{enumerate}\vspace{-1.5ex} 
Assuming that the phase modulation is approximately linear unless the carrier is strong enough to affect the dc bias, two basic rules hold:\vspace{-1.5ex} 
\begin{enumerate} 
\item $b_{-1}$ is independent of the carrier power. 
\item Cascading two or more devices the $b_{-1}$ add up, regardless of the device order in the chain.
\end{enumerate}\vspace{-1.5ex} 
The reader may have in mind the Friis formula for the noise referred to the input of a chain \cite{friis44ire}, stating that the contribution of each stage is divided by the total gain of the preceding stages, and therefore indicating the first stage as the major noise source.  The Friis formula raises from the additive property of white noise, hence it does not apply to parametric noise. 

Early measurements on \emph{amplifiers} \cite{halford68fcs,walls97uffc,hati03fcs} suggest that different amplifiers based on a given technology tend to have about
the same $b_{-1}$, and that $b_{-1}$ is nearly constant in a wide range of carrier frequency and power.
Our independent experiments confirm that the $1/f$ phase noise of a given amplifier is independent of power in a wide range \cite{rubiola08arxiv-amplifier-noise}, \cite{boudot:phd}, \cite[Chapter 2]{leeson-cambridge}.  
For example, $b_{-1}$ of a commercial amplifier (Microwave Solutions
MSH6545502)
    %\footnote{Vol.~7 p.~135}
that we measured at 9.9 GHz is between $1.25{\times}10^{-11}$ and
$2{\times}10^{-11}$ from 300 $\mu$W to 80 mW of output power.  Similarly, the $1/f$ noise of a LNPT32 SiGe amplifier measured between 32 $\mu$W ($-15$ dBm) and 1 mW (0 dBm) in 5 dB steps overlap perfectly.  
In summary, the typical phase flickering $b_{-1}$ of a ``good'' microwave amplifier is between $10^{-10}$ and $10^{-12}$ \unit{rad^2/Hz} ($-100$ to $-120$ \unit{dBrad^2/Hz}) for the GaAs HBTs, and between $10^{-12}$ and $10^{-13}$ \unit{rad^2/Hz} ($-120$ to $-130$ \unit{dBrad^2/Hz}) for the SiGe transistors. 

The $1/f$ noise of the \emph{microwave photodetector} is expected to be similar to that of an amplifier because the underlying physics and technology are similar.  The measurement is a challenging experimental problem, which has been tackled only at the JPL independently by Rubiola and Shieh \cite{maleki98ofc,shieh05ptl,rubiola06mtt-photodiodes}.  The results agree in that at 10 GHz the typical $b_{-1}$ of a InGaAs p-i-n photodetector is of $10^{-12}$ \unit{rad^2/Hz} ($-120$ \unit{dBrad^2/Hz}).

Microwave \emph{variable attenuators} and \emph{variable phase shifters } can be necessary for adjustment.  Our early measurements \cite{rubiola99rsi} indicate that the $b_{-1}$ of these components is of the order of $10^{-15}$ \unit{rad^2/Hz} ($-150$ \unit{dBrad^2/Hz}), which is negligible as compared to the amplifiers and to the photodetectors.

Additional sources of noise are the EOM, the laser amplified spontaneous emission, and the noise of the optical pump.  As theory provide no indications about these effects, a pragmatic approach is necessary, which consists of measuring the total noise of the microwave delay unit in different configurations.

\section{Dual-delay phase noise measurement}\label{sec:dual-channel}
%------------------------------------------------------------------------
\begin{figure*}[t]
\centering\namedgraphics{0.7}{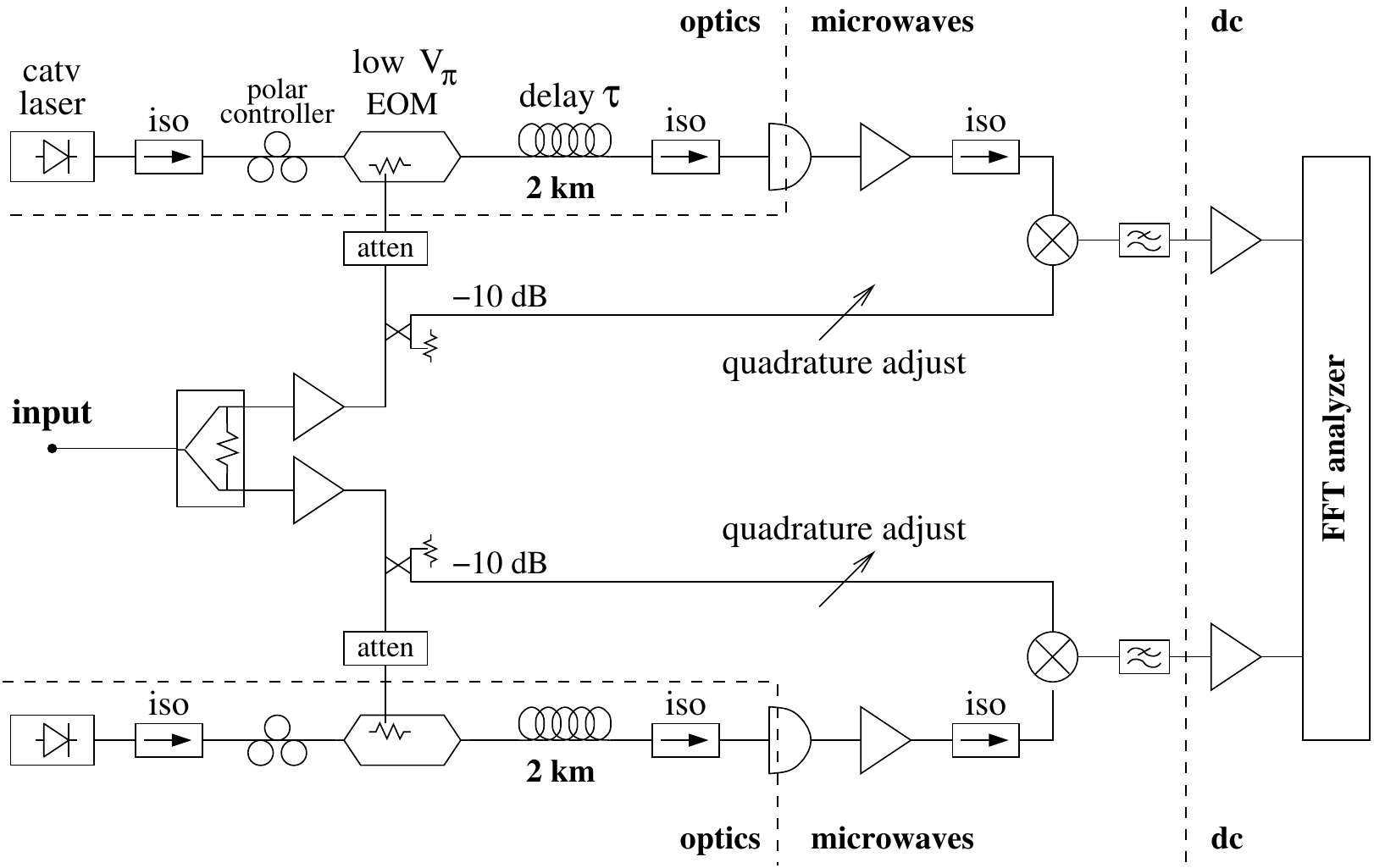}{\textwidth}
\caption{Scheme of the dual-channel instrument.}
\label{fig:ddl-correl-instrum}
\end{figure*}
Figure~\ref{fig:ddl-correl-instrum} shows the scheme of the instrument.
However similar to the previous ones (\cite{salik04fcs-xhomodyne,salzenstein07appa-dual-delay-line}), this version adds engineering and a substantial progress in understanding $1/f$ noise.

The instrument consists of two equal and independent channels that measure the oscillator under test using the principle of Fig.~\ref{fig:ddl-delay-basic}.  Then, the single-channel noise is removed using the cross-spectrum method before converting the mixer output into the oscillator noise $S_\varphi(f)$ with Eq.~\req{eqn:ddl-hphi2}.

Looking at one channel, we observe that the microwave signal is split into two branches before the EOM, so that the long branch consists of modulator, optical fiber (delay $\tau$), photodetector and microwave amplifier, while the short branch is a microwave path of negligible length.  This differs from the first single-channel instrument \cite[Fig.\,7]{rubiola05josab-delay-line}, in which the short branch was optical.  Removing the photodetector and the microwave amplifier from the short branch yields lower noise, and in turn faster convergence of the correlation algorithm.   Lower laser power is needed.  Another advantage is that we use a microwave power splitter instead of an optical power splitter.  
While the noise of the former is negligible for our purposes \cite{rubiola00rsi-correlation,rubiola02rsi-matrix}, we have no first-hand knowledge about the latter.   The problem with this configuration is that we no longer have an optical input, so the microwave-modulated light beams can not be measured.

Trading off with the available components, we had to use both GaAs amplifiers and SiGe amplifiers.  Since the SiGe amplifiers exhibit lower $1/f$ noise, they are inserted at the photodetector output.    
This choice is motivated by the fact that the $1/f$ noise at the mixer input is converted into oscillator $1/f^3$ noise by Eq.~\req{eqn:ddl-sphi-o}--\req{eqn:ddl-hphi2} while the $1/f$ noise at the EOM input is not.

\subsection{Mixer Noise}\label{ssec:ddl-mixer-noise}
%--------------------------------------------------------------------
Used as a phase detector, the double-balanced mixer needs to be saturated at both inputs.  The conversion gain is of 0.1--0.5 V/rad when  the power is in the appropriate range, which is of ${\pm}5$ dB centered around the optimum power of 5--10 mW\@. 
Out of this range, $b_{-1}$ increases.  At lower power the conversion gain drops suddenly because the input voltage is insufficient for the internal Schottky diodes to switch.

Out of experience in low-flicker applications, we have a preference for the mixers manufactured by Narda and by Marki.  The coefficient $b_{-1}$ is of the order of $10^{-12}$, similar to that of the photodetectors.
The white noise is chiefly the noise of the output amplifier
divided by the conversion gain $k_\varphi$.  Assuming that the amplifier noise is 1.6 \unit{nV/\sqrt{Hz}} (our low-flicker amplifiers input-terminated to 50 \ohm\ \cite{rubiola04rsi-amplifier}) and that $k_\varphi=0.1$ V/rad (conservative with respect to $P_0$), the white noise is $b_0=2.5{\times}10^{-16}$ \unit{rad^2/Hz} ($-156$ \unit{dBrad^2/Hz}).

Mixers are sensitive to the amplitude noise $\alpha(t)$ of the input signal, hence the
mixer output takes the form $v(t)=k_\varphi\varphi(t)+k_\alpha\alpha(t)$.  This is due to the asymmetry of the internal diodes and baluns.  In some cases we have measured $k_\varphi/k_\alpha$ of 5 or less, while values of 10--20 are common. 
In photonic systems the contamination from amplitude noise can be a serious problem because of the power of some lasers and laser amplifiers fluctuates.  Brendel \cite{brendel77im} and Cibel \cite{cibiel02uffc} suggest that the mixer can be operated at a sweet point off the quadrature, where the sensitivity AM noise nulls.  A further study shows that the Brendel offset method can not be used in our case \cite{rubiola07uffc-am-to-pm-pollution} because the null of amplitude sensitivity results from the equilibrium between equal and opposite sensitivities at the two inputs.  But the delay de-correlates the mixer input signals by virtue of the same mechanism used to measure $S_\varphi(f)$.

\subsection{Calibration}
%-------------------------------------------------------
The phase-noise measurement system is governed by $S_{v}(f)=k_\varphi^2 |H(jf)|^2S_{\varphi}(f)$ [Eq.~\req{eqn:ddl-sphi-o} and \req{eqn:ddl-hphi2}].  Calibration takes the accurate measurement of $\tau$ and $k_\varphi$.  An accuracy of 1 dB can be expected.  Since the stability of the optical fiber exceeds our needs, $\tau$ does not need re-calibration after first set-up.  Optical reflection methods are suitable, as well as the spectrum measurement of the noise the peak at $f=1/\tau$, where $|H(jf)|^2=0$.  Conversely, $k_\varphi$ is highly sensitive to optical and microwave power.  It is therefore recommended to measure it at least every time the experimental conditions are changed.

The simplest way to measure $k_\varphi$ is to introduce a reference sinusoidal modulation $\varphi(t)=m_\varphi\sin2\pi f_mt$ in the microwave signal \req{eqn:ddl-clock-signal}, after replacing the oscillator under test with a microwave synthesizer.  This accounts for the dc amplifier at the mixer output, not shown in Fig.~\ref{fig:ddl-delay-basic}.  It is convenient to set $f_m\lesssim0.1/\tau$, so that it holds $\sin(\pi f_m\tau)\simeq\pi f_m\tau$ in \req{eqn:ddl-hphi2}.  With commercial synthesizers, frequency modulation is more suitable then phase modulation because the frequency deviation $\Delta f$ can be measured in static conditions, with dc input.  Frequency modulation is equivalent to phase modulation
\begin{align}
\varphi(t)=\frac{\Delta f}{f_m}\sin2\pi f_mt
\label{eqn:ddl-reference-fm}
\end{align}
of index $m_\varphi=\Delta f/f_m$.
Substituting \req{eqn:ddl-reference-fm} in \req{eqn:ddl-clock-signal} and truncating the series expansion to the first-order, the microwave signal is
\begin{align}
v(t)&=J_0\left(\tfrac{\Delta f}{f_m}\right)\cos2\pi\nu_0t
+ J_1\left(\tfrac{\Delta f}{f_m}\right)
\bigl[\cos2\pi(\nu_0+f_m)t-\cos2\pi(\nu_0-f_m)t \bigr]~,
\label{eqn:ddl-reference-signal-a}
\end{align}
where $J_n(~)$ is the Bessel function of order $n$.  For small $m_\varphi$, we use the approximations $J_0(m_\varphi)\simeq1$ and $J_1(m_\varphi)\simeq\frac12m_\varphi$.  Thus, 
\begin{align}
v(t)&=\cos2\pi\nu_0t
+ \frac12 \frac{\Delta f}{f_m}
\bigl[\cos2\pi(\nu_0+f_m)t-\cos2\pi(\nu_0-f_m)t \bigr]~.
\label{eqn:ddl-reference-signal-b}
\end{align}
Finally, the modulation index can be easily calibrated by inspection with a microwave spectrum analyzer.  In this case it is recommended to measure $m_\varphi$ with relatively large sidebands (say, 40 dB below the carrier), and then to reduce $m_\varphi$ by inserting a calibrated attenuator at the FM input of the synthesizer.  This enhances the accuracy of the spectrum analyzer.  For example, having $\tau=10$ $\mu$s and $k_\varphi=0.2$ V/rad, we may set $f_m=5$ kHz and $m_\varphi=2{\times}10^{-2}$ rad.
In this case, the detected signal has a peak voltage of 400 $\mu$V at the mixer output, thus 40 mV after 40 dB amplification.  Measuring $k_\varphi$, it may be convenient to attenuate the modulating signal by 20--30 dB, so that the system is calibrated in actual operating conditions.

\section{Experiments}
%------------------------------------------------------------------------

\subsection{Measurement of the background at zero fiber length}
%-------------------------------------------------------
\begin{figure}
\centering\namedgraphics{0.60}{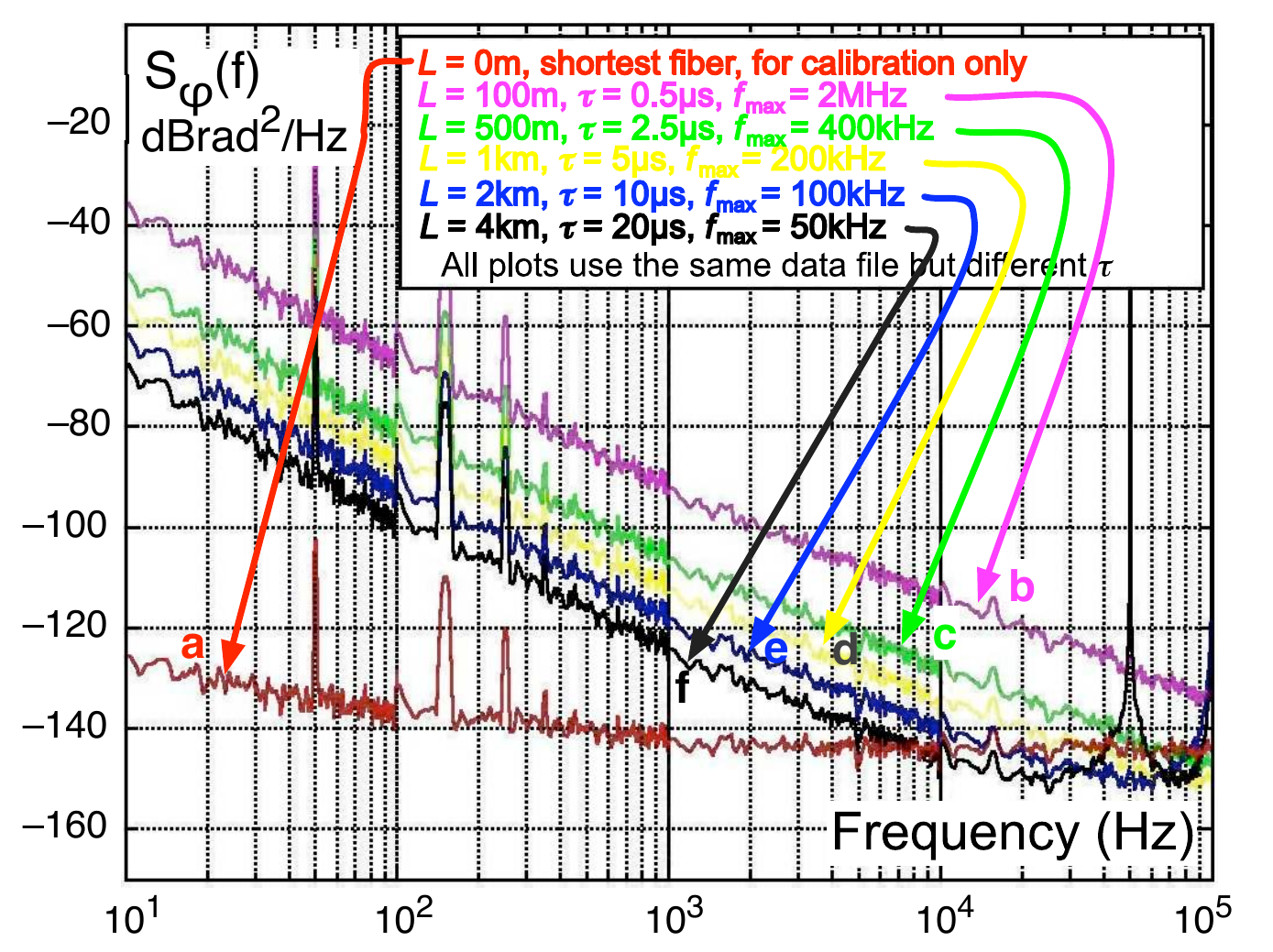}{\columnwidth}
\caption{Measured single-channel background noise with zero-length optical fiber.}
\label{fig:ddl-bground-zero-fber}
\end{figure}
Replacing the spools of Fig.~\ref{fig:ddl-correl-instrum} with short fibers, the oscillator phase noise is rejected.  The noise phenomena originated inside the fiber, or taken in by the fiber, are also eliminated.  The noise measured in this conditions (Fig.~\ref{fig:ddl-bground-zero-fber}) is the instrument background as it would be with noise-free fibers.  The curve \textsf{a} (red) is the phase noise $S_\theta(f)$ measured the mixer output.
The other curves (\textsf{b}--\textsf{f}) are plotted using the same data set, after unapplying  $k_\varphi^2|H(jf)|^2$ for various fiber lengths [Eq.~\req{eqn:ddl-sphi-o}--\req{eqn:ddl-hphi2}].  As expected after Sec.~\ref{ssec:ddl-delay-line-theory},
the curve f (black, 4 km fiber, $\tau=20$ $\mu$s) shows a peak at $1/\tau=50$ kHz and at $2/\tau=100$ kHz, where $|H(jf)|^2=0$.   Additionally, the curve \textsf{f} shows two minima 6 dB lower than $S_\theta(f)$ at $f=25$ kHz and at $f=75$ kHz, where $|H(jf)|^2=4$.  The same facts are observed at twice the frequency on the curve \textsf{e} (blue, 2 km fiber, $\tau=10$ $\mu$s).

\subsection{Effect of microwave AM noise and of laser RIN}
%-------------------------------------------------------
\begin{figure}
\centering\namedgraphics{0.65}{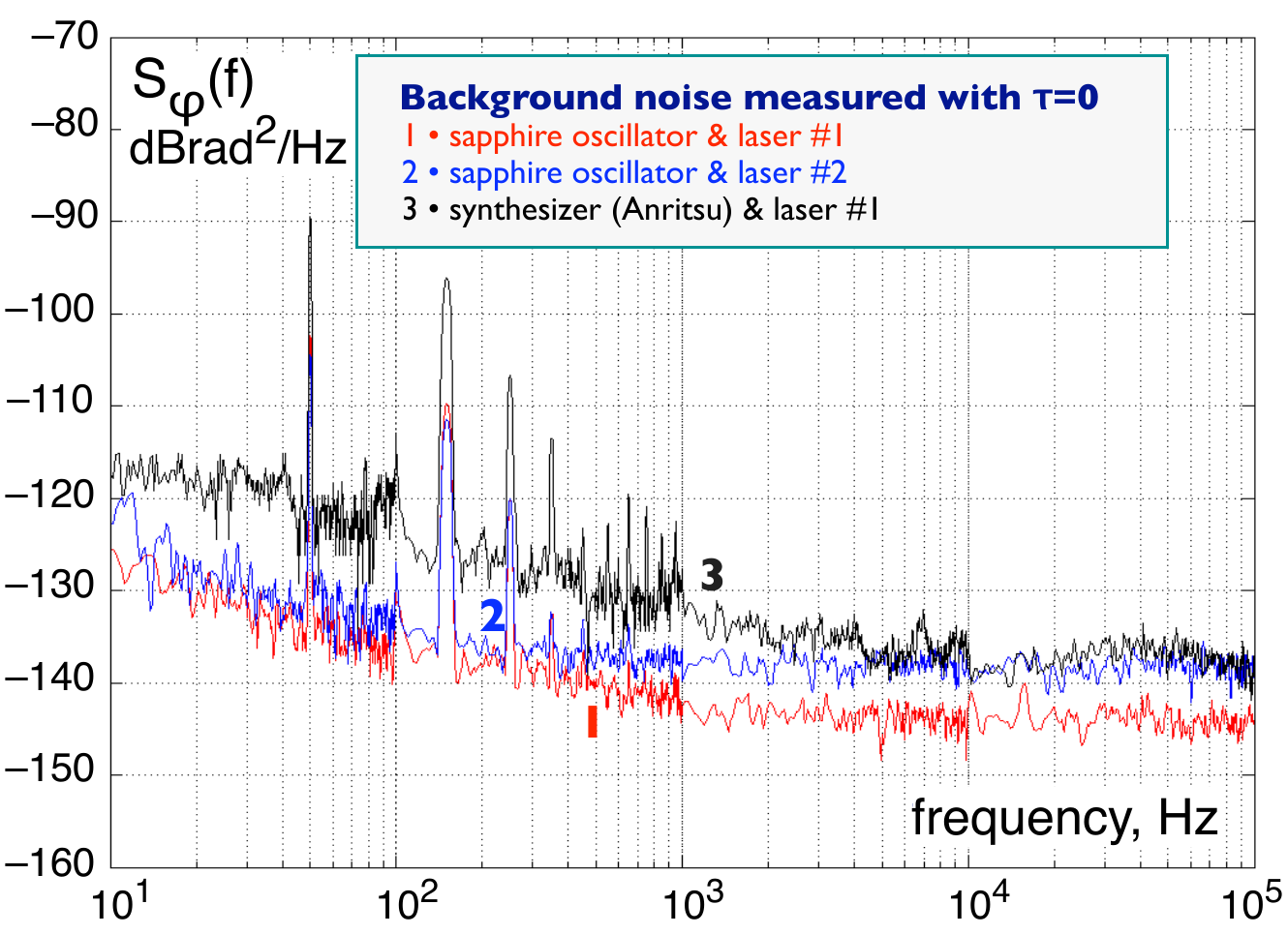}{\columnwidth}
\caption{Effect of the laser RIN and of the oscillator-under-test AM noise, measured with zero-length fiber.}
\label{fig:ddl-effect-of-am}
\end{figure}
This experiment shows qualitatively the effect of the laser RIN and of the AM noise of the oscillator under test, still in single-channel mode and with zero-length optical fiber, so that the oscillator phase noise is rejected.  The curve \textsf{3}  (red) is the same as the curve \textsf{a} of Fig.~\ref{fig:ddl-bground-zero-fber}.  In this case, the optical source is a CATV laser, while the oscillator under test is a 10 GHz sapphire-loaded dielectric-cavity oscillator operated at room temperature \cite{giordano05epjap-oscillators}.  Besides high stability, the sapphire oscillator performs low AM noise.  Replacing the laser with a different one with higher RIN (curve \textsf{2}, blue), or replacing the oscillator under test with a synthesizer (curve \textsf{3}, black), which has higher AM noise, the background increases significantly.

\subsection{Measurement of a sapphire microwave oscillator}
%-------------------------------------------------------
\begin{figure}
\centering\namedgraphics{0.6}{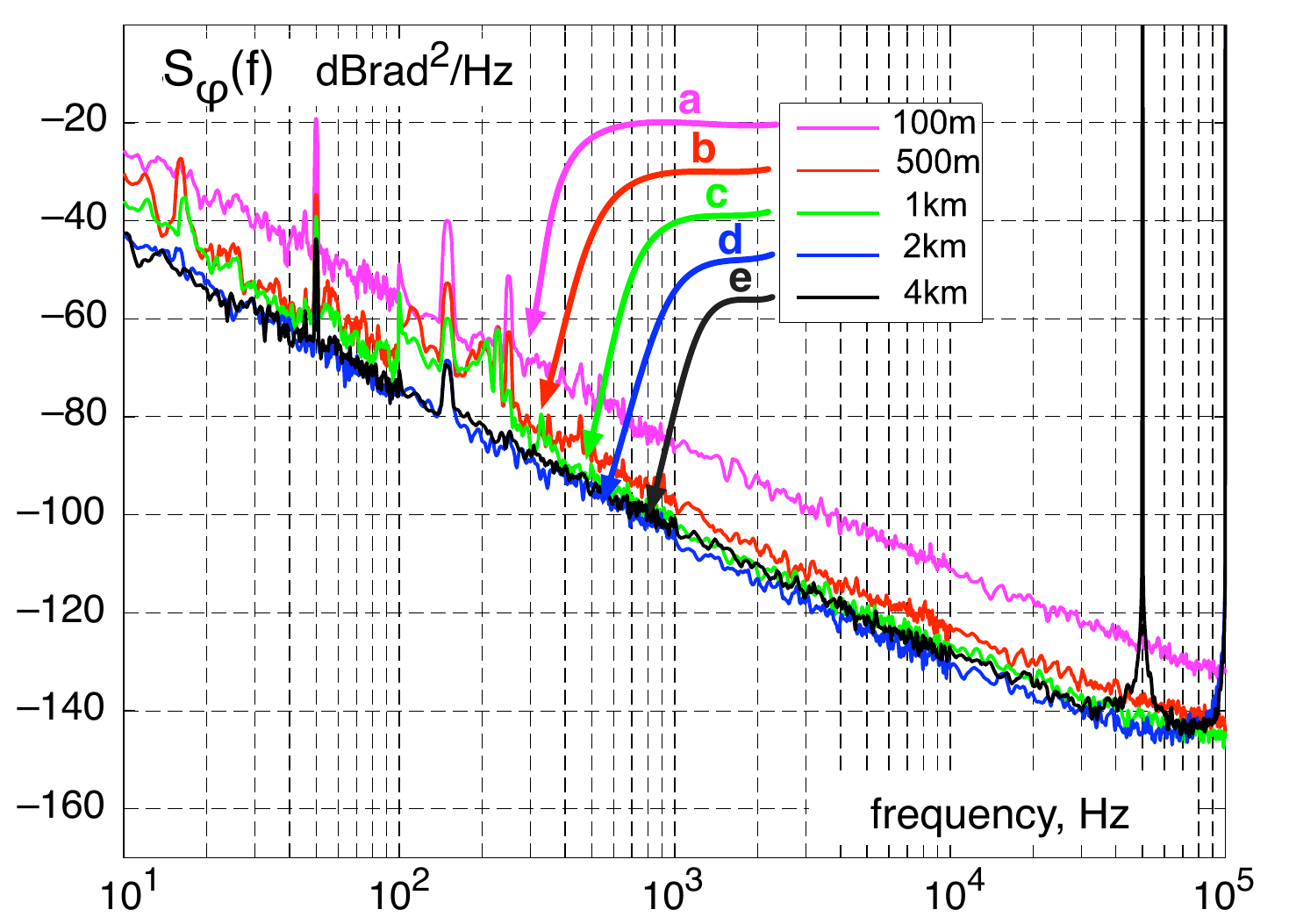}{\columnwidth}
\vspace*{-2ex}
\caption{Phase noise of a sapphire oscillator.}
\label{fig:ddl-saphire-noise-spectrum}
\end{figure}
We measured the phase-noise of a room-temperature sapphire oscillator, still in single-channel mode, progressively increasing the fiber length.  Let us focus on the $1/f^3$ noise, which dominates on the spectrum  (Fig.~\ref{fig:ddl-saphire-noise-spectrum}).  When the delay is insufficient to detect the noise of the source under test, the spectrum is the background of the instrument, which scales with the inverse square length, that is, $-6$ dB in $S_\varphi(f)$ for a factor of 2 in the delay.  This is visible on the curves \textsf{a} (magenta, 100 m, $\tau=0.5$ $\mu$s) and \textsf{b} (red, 500 m, $\tau=2.5$ $\mu$s).  Increasing the length, the  $1/f^3$ noise no longer decreases.  This fact, seen on the curves \textsf{d} (blue, 2 km, $\tau=10$ $\mu$s) and \textsf{e} (black, 4 km, $\tau=20$ $\mu$s), indicates that the instrument measures the phase noise of the sapphire oscillator, which does not scale with the delay.
Measuring the $1/f^3$ noise of a sapphire oscillator in single-channel mode is a remarkable result because this oscillator is regarded as the best reference in the field of low-noise microwave sources.

\subsection{Assessing the noise of the photonic channel}
%-------------------------------------------------------
\begin{figure*}[t]
\centering\namedgraphics{0.70}{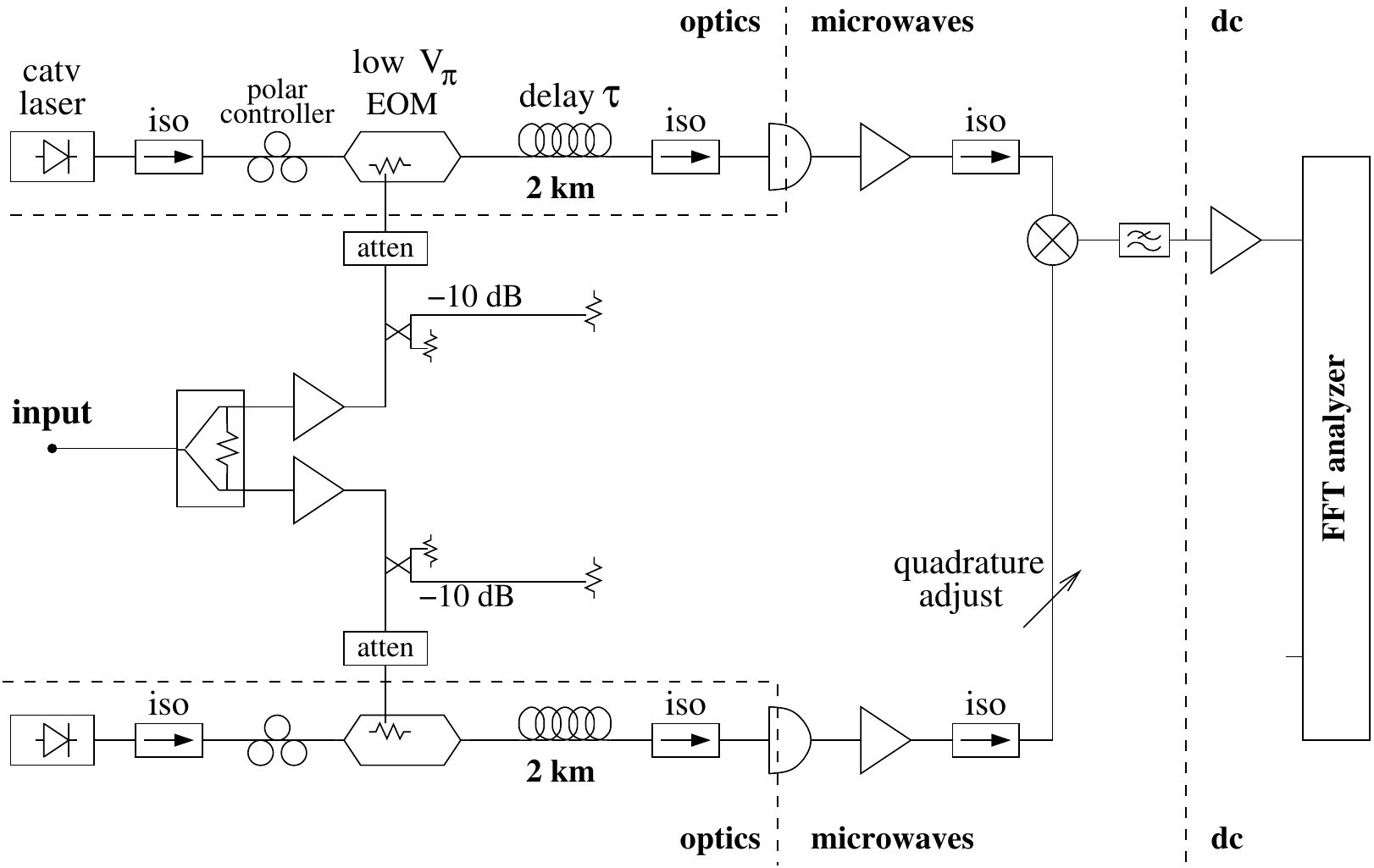}{\textwidth}
\caption{Measurement of the background noise, including the optical fibers.}
\label{fig:ddl-bground-noise-scheme}
\end{figure*}
\begin{figure}
\centering\namedgraphics{0.6}{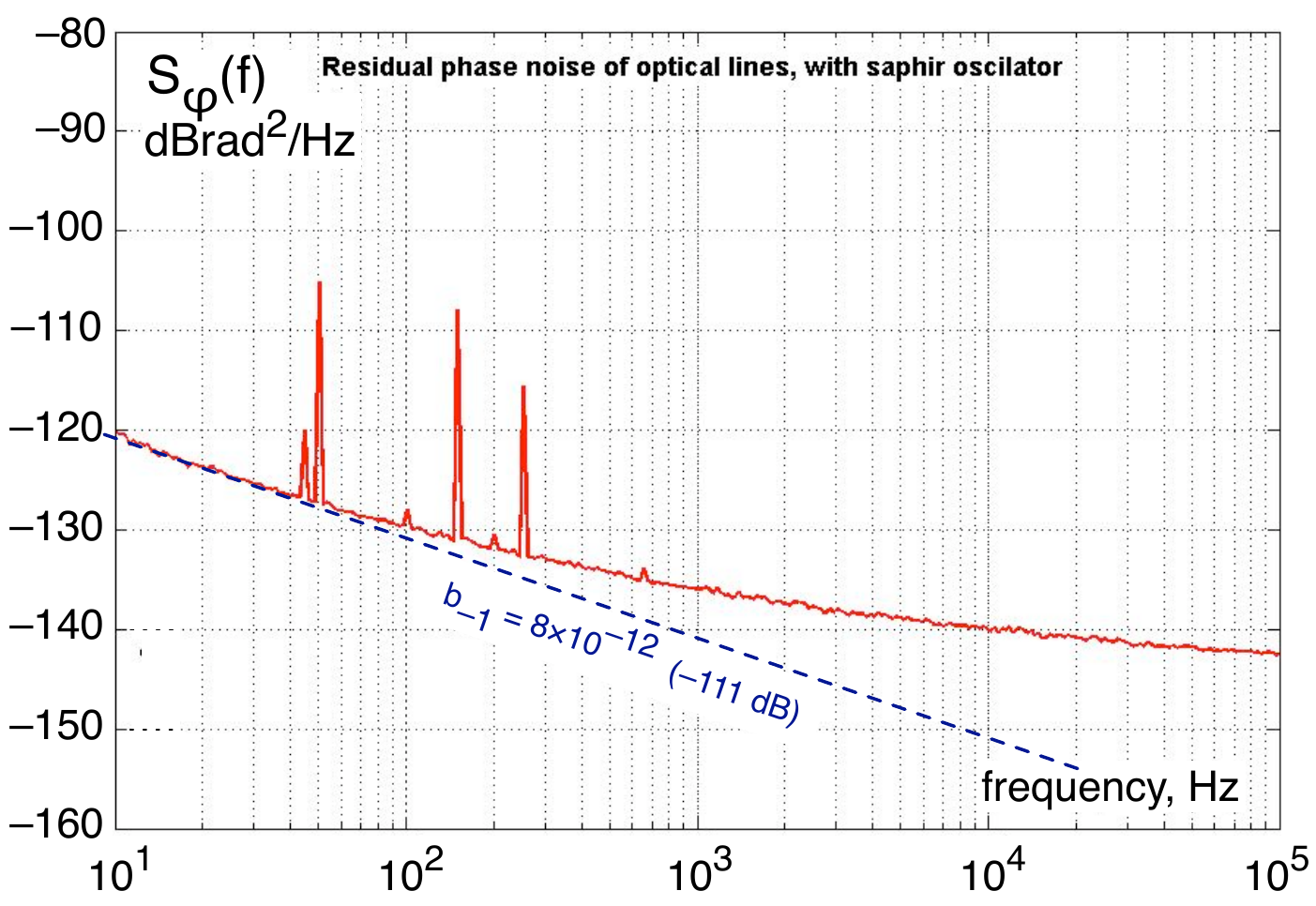}{\columnwidth}
\vspace*{-2ex}
\caption{Background noise, including the optical fibers.}
\label{fig:ddl-bground-noise-spectrum}
\end{figure}
We measured the phase noise of the photonic channel using the scheme of Fig.~\ref{fig:ddl-bground-noise-scheme}, derived from Fig.~\ref{fig:ddl-correl-instrum} after removing some parts.  The phase noise of the reference oscillator is rejected by using two fibers of equal length.   This experiments suffers from the following limitations.\vspace{-1.5ex}
\begin{enumerate}\addtolength{\itemsep}{-1.5ex}
\item The noise of the fibers can not be separated from other noises.
\item The $1/f$ noise of the GaAs amplifiers that drive the EOM can show up.  
\item We could not use the correlation method because it takes four matched optical delay-lines, which were not available.  
\end{enumerate}\vspace{-1.5ex}
Nonetheless, this scheme has the merit of giving at least an upper bound of the achievable noise.  The measured spectrum, shown in Fig.~\ref{fig:ddl-bground-noise-spectrum}, indicates that the $1/f$ phase noise is $b_{-1}=8{\times}10^{-12}$ \unit{rad^2/Hz} ($-111$ \unit{rad^2/Hz}).  At this level, the mixer noise is negligible.

\subsection{Background noise of the two-channel instrument}
%-------------------------------------------------------
\begin{figure}[t]
\centering\namedgraphics{0.70}{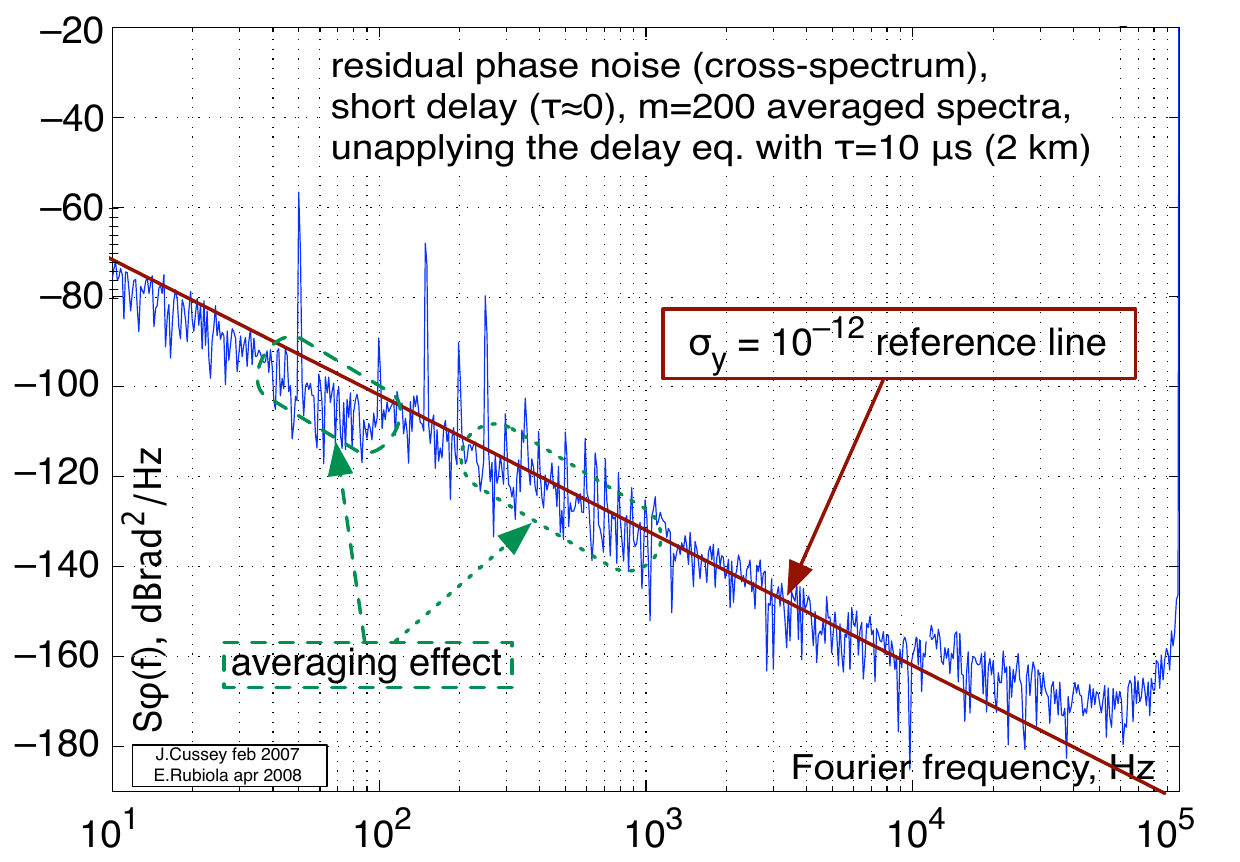}{\columnwidth}
\caption{Background noise in two-channel mode, measured with zero-length fiber.}
\label{fig:ddl-bground-zero-fiber-correl}
\end{figure}
We measured the background noise in the two-channel configuration, using the cross-spectrum method of Sec.~\ref{ssec:ddl-xspectrum} and with zero-length optical fiber, so that the phase noise of the 10 GHz reference oscillator is rejected.  This experiment does not account for the optical noises originated inside the fibers, these phenomena are rejected in Eq.~\req{eqn:ddl-Syx-avg} because the two fibers can not be correlated.  When this experiment was done, the stability of the quadrature condition was still insufficient for long acquisitions.  For this reason we stopped the measurement after $m=200$ spectra.  The cross spectrum is shown in Fig.~\ref{fig:ddl-bground-zero-fiber-correl}.

The reference straight line (red) is the $1/f^3$ phase noise equivalent to the Allan deviation $\sigma_y=10^{-12}$, calculated with Eq.~\req{eqn:ddl-sy} and \req{eqn:ddl-sy-sigmay}.  Thus, averaging on 200 spectra the instrument can measure the stability of an oscillator at $10^{-12}$ level.

The signature of a correlated noise is a smooth cross spectrum.  Yet, in Fig.~\ref{fig:ddl-bground-zero-fiber-correl} the variance is far too large for the two channels to be correlated.  This indicates that the sensitivity is limited by the small number of averaged spectra.  Additionally, the spectrum shows a sawtooth-like discrepancy vs.\ the $1/f^3$ line.  This averaging effect, due to the measurement bandwidth that increases with frequency in logarithmic resolution, further indicates that the value of $m$ is still insufficient.  The ultimate sensitivity for large $m$ is still not known.

\subsection{Opto-electronic oscillator}\label{ssec:oeo-prototype}
%-------------------------------------------------------
\begin{figure}[t]
\centering\namedgraphics{0.22}{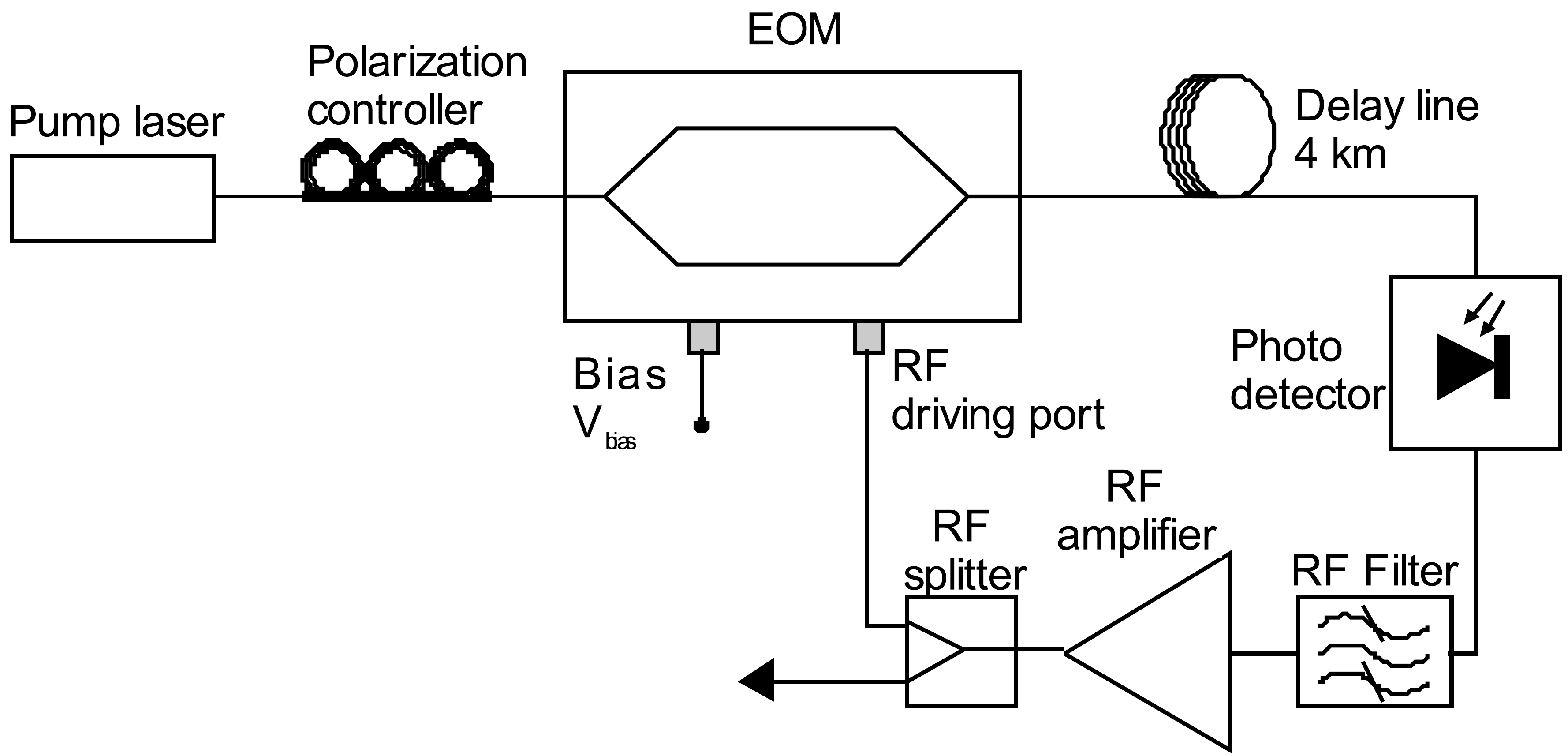}{\columnwidth}
\caption{Scheme of the opto-electronic oscillator.}
\label{fig:ddl-oeo-scheme}
\end{figure}
\begin{figure}[t]
\centering\namedgraphics{0.25}{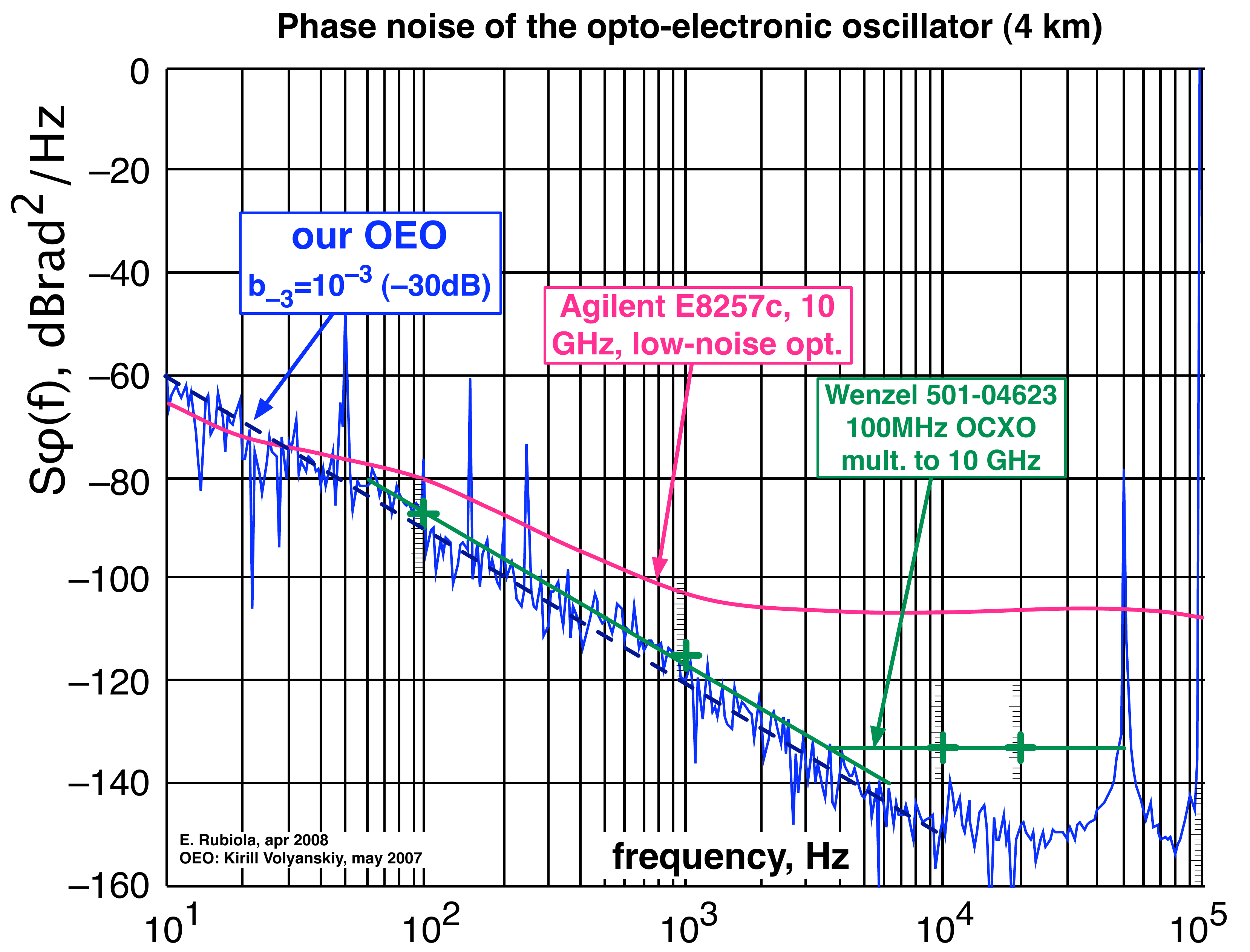}{\columnwidth}
\vspace*{-3ex}
\caption{Phase noise of the opto-electronic oscillator.}
\label{fig:ddl-oeo-phase-noise}
\end{figure}
We implemented the oscillator of Fig.~\ref{fig:ddl-oeo-scheme} using a 4 km delay line (20 $\mu$s), a SiGe amplifier, and a photodetector with integrated transconductance amplifier.  Oscillation starts at 6--7 mW optical power.  The best working point occurs at 12--13 mW optical power, where the microwave output power is 20 mW\@.  For flicker noise the microwave chain is similar to one channel in Fig.~\ref{fig:ddl-bground-noise-scheme} because the order of the devices in the chain is not relevant.
Thus, we take $b_{-1}=8{\times}10^{-12}$ \unit{rad^2/Hz} ($-111$ \unit{dBrad^2/Hz}), the same of Fig.~\ref{fig:ddl-bground-noise-spectrum}, as the first estimate of the loop noise.  Though somewhat arbitrary, this value accounts for the photodetector internal amplifier, more noisy than our microwave amplifiers.  Using $(b_{-1})_\text{loop}=8{\times}10^{-12}$ \unit{rad^2/Hz} in the oscillator noise model of Section~\ref{ssec:ddl-oscillator-noise}, the expected oscillator flickering is $(b_{-3})_\text{osc}=6.3{\times}10^{-4}$ \unit{rad^2/Hz} ($-32$ \unit{dBrad^2/Hz}).  By virtue of \req{eqn:ddl-sy} and \req{eqn:ddl-sy-sigmay}, this is equivalent to a frequency stability $\sigma_y=2.9{\times}10^{-12}$ (Allan deviation).
The noise spectrum (Fig~\ref{fig:ddl-oeo-phase-noise}), measured with the dual-channel instrument, shows a frequency flicker of $10^{-3}$ \unit{rad^2/Hz} ($-30$ \unit{dBrad^2/Hz}), equivalent to $\sigma_y=3.6{\times}10^{-12}$.  The discrepancy between predicted value and result is of 2 dB\@.

Interestingly, the OEO phase noise compares favorably to the lowest-noise microwave synthesizers and quartz oscillator multiplied to 10 GHz.  Yet, the OEO can be switched in steps of 50 kHz without degrading the noise.

\section{Further developments}
Our experiments suggest the following improvements, tested or in progress. 
\begin{enumerate}

\item In the phase-noise instrument, the microwave power the input of the EOM and the two inputs of the mixer is a critical parameter because the gain of the entire system is strongly non-linear.  The insertion of test points is recommended, using common and inexpensive power detectors after tapping a fraction of the power with 20 dB directional couplers.

\item The instrument requires that the two signals at the mixer input are kept in quadrature.  This is usually accomplished with a line stretcher or a with  voltage-controlled phase shifter.  A smarter solution exploits the dispersion of the optical fiber, by adjusting the laser wavelength via the temperature control \cite{laser-tuning}.  This is viable only at 1.55 $\mu$m and with long fibers.  For reference, using a 2 km fiber and our lasers, it takes 10 K of temperature change for a quarter wavelength of the microwave signal.  Another solution exploits the temperature coefficient of the fiber delay, $6.85{\times}10^{-6}$/K\@.  Accordingly, a quarter wavelength takes 0.37 K of temperature with 2 km fiber (10 $\mu$s), or 3.7 K with 200 m.

\item Some popular EOMs have a low-frequency photodetector at the unused output port of the Mach-Zehnder interferometer, which we hope to exploit to stabilize the bias point.

\item The OEO frequency can be fine-tuned by adding a RF signal with an SSB modulator at the output.  Extremely high resolution, of the order of  $10^{-16}$ can be obtained with a 48 bit DDS, with no degradation of the spectral purity \cite{elisa,analog-devices-AD9854}.

\end{enumerate}

\subsection*{Acknowledgments}
We are indebted with Lute Maleki (OEwaves and NASA/Caltech JPL, Pasadena, CA, USA),  Nan Yu (NASA/Caltech JPL) and Ertan Salik (NASA/Caltech JPL, now with the California State Polytechnic University, Pomona, CA, USA) for having taught us a lot in this domain, and for numerous discussions, advices and exchanges.  Gilles Cibiel (CNES, France) provided contract support and help.  Rodolphe Boudot (FEMTO-ST, now with SYRTE, Paris, FRANCE) helped with the sapphire oscillator,  Xavier Jouvenceau (FEMTO-ST) helped in the implementation of the early version of the correlation system, and Cyrus Rocher (FEMTO-ST) implemented the temperature control.  

This work is supported with grants from Aeroflex, CNES and ANR.

%==================================
%\def\bibfile#1{/Volumes/rubiola/Documents/articles/bibliography/#1}
\def\bibfile#1{/Users/rubiola/Documents/articles/bibliography/#1}
\addcontentsline{toc}{section}{References}
\bibliographystyle{abbrv}
\bibliography{\bibfile{ref-short},\bibfile{references},\bibfile{rubiola},ddl-local}

\begin{thebibliography}{10}

\bibitem{analog-devices-AD9854}
{A}nalog {D}evices {AD9854} {DDS}.
\newblock http://www.analog.com/.

\bibitem{boudot:phd}
R.~Boudot.
\newblock {\em Oscillateurs micro-onde \`{a} haute puret\'{e} spectrale}.
\newblock {Ph.\,D}. thesis, Universit\'{e} de Franche Comt\'{e}, Besan\c{c}on,
  France, Dec.~7, 2006.

\bibitem{brendel77im}
R.~Brendel, G.~Marianneau, and J.~Ubersfeld.
\newblock Phase and amplitude modulation effects in a phase detector using an
  incorrectly balanced mixer.
\newblock {\em IEEE Trans.\ Instrum.\ Meas.}, 26(2):98--102, June 1977.

\bibitem{ccir90rep580-3}
{CCIR Study Group VII}.
\newblock Characterization of frequency and phase noise, {R}eport no.\ 580-3.
\newblock In {\em Standard Frequencies and Time Signals}, volume {VII} (annex)
  of {\em Recommendations and Reports of the {CCIR}}, pages 160--171.
  International Telecommunication Union {(ITU)}, Geneva, Switzerland, 1990.

\bibitem{chang:rf-photonics}
W.~S.~C. Chang, editor.
\newblock {\em {RF} Photonic Technology in Optical Fiber Links}.
\newblock Cambridge, Cambridge, UK, 2002.

\bibitem{chembo08jqe-oeo-phase-diffusion}
Y.~K. Chembo, K.~Volyanskiy, L.~Larger, E.~Rubiola, and P.~Colet.
\newblock Determination of phase noise spectra in optoelectronic microwave
  oscillators: a phase diffusion approach.
\newblock {\em J. Quantum Electron.}, pages **--**, 2008.
\newblock Accepted for publication, July 2008. Also arXiv:0805.3317v1
  [physics.optics], May 2008.

\bibitem{cibiel02uffc}
G.~Cibiel, M.~R\'{e}gis, E.~Tournier, and O.~Llopis.
\newblock {AM} noise impact on low level phase noise measurements.
\newblock {\em IEEE Trans.\ Ultras.\ Ferroelec.\ and Freq.\ Contr.},
  49(6):784--788, June 2002.

\bibitem{ye03rmp}
S.~T. Cundiff and J.~Ye.
\newblock Colloquium: Femtosecond optical frequency combs.
\newblock {\em Rev.\ Mod.\ Phys.}, 75(1):325--342, Jan. 2003.

\bibitem{elisa}
Elisa - technical notes of the esa cryo project.
\newblock Series of FEMTO-ST and ESA internal reports, 2007-08.

\bibitem{eliyahu03fcs}
D.~Eliyahu and L.~Maleki.
\newblock Low phase noise and spurious level in multi-loop opto-electronic
  oscillators.
\newblock In {\em Proc. Europ.\ Freq.\ Time Forum and Freq.\ Control Symp.
  Joint Meeting}, pages 405--410, Tampa (FL, USA), May~5--8, 2003.

\bibitem{friis44ire}
H.~T. Friis.
\newblock Noise figure of radio receivers.
\newblock {\em Proc.\ IRE}, 32:419--422, July 1944.

\bibitem{giordano05epjap-oscillators}
V.~Giordano, P.-Y. Bourgeois, Y.~Gruson, N.~Boubekeur, R.~Boudot, E.~Rubiola,
  N.~Bazin, and Y.~Kersal\'{e}.
\newblock New advances in ultra-stable microwave oscillators.
\newblock {\em Euro.\ Phys.\ J. -- Appl.\ Phys.}, 32(2):133--141, Nov. 2005.

\bibitem{halford68fcs}
D.~Halford, A.~E. Wainwright, and J.~A. Barnes.
\newblock Flicker noise of phase in {RF} amplifiers: Characterization, cause,
  and cure.
\newblock In {\em Proc.\ Freq.\ Control Symp.}, pages 340--341, Apr.~22--24,
  1968.
\newblock Abstract only is published.

\bibitem{hati03fcs}
A.~Hati, D.~Howe, D.~Walker, and F.~Walls.
\newblock Noise figure vs.\ {PM} noise measurements: A study at microwave
  frequencies.
\newblock In {\em Proc. Europ.\ Freq.\ Time Forum and Freq.\ Control Symp.
  Joint Meeting}, Tampa, {FL}, May~5--8, 2003.

\bibitem{ippen03ol}
D.~J. Jones, K.~W. Holman, M.~Notcutt, J.~Ye, J.~Chandalia, L.~A. Jiang, E.~P.
  Ippen, and H.~Yokoyama.
\newblock Ultralow-jitter, 1550-nm mode-locked semiconductor laser synchronized
  to a visible optical frequency standard.
\newblock {\em Optics Lett.}, 28(10):813--815, May~15 2003.

\bibitem{kimball:precise-frequency}
H.~G. Kimball, editor.
\newblock {\em Handbook of selection and use of precise frequency and time
  systems}.
\newblock ITU, 1997.

\bibitem{kippenberg04prl}
T.~J. Kippenberg, S.~M. Spillane, and K.~J. Vahala.
\newblock Kerr-nonlinearity optical parametric oscillation in an ultrahigh-{Q}
  toroid microcavity.
\newblock {\em Phys.\ Rev.\ Lett.}, 93(8):083904--1--083904--4, Aug.~20, 2004.

\bibitem{leeson-arxiv}
E.~Rubiola.
\newblock The {L}eeson effect.
\newblock ar{X}iv:physics/0502143v1, web site arxiv.org, Feb. 2005.
\newblock Abridged draft version of \cite{leeson-cambridge}.

\bibitem{leeson-cambridge}
E.~Rubiola.
\newblock {\em Phase Noise and Frequency Stability in Oscillators}.
\newblock Cambridge University Press, Cambridge, MA, Nov. 2008.
\newblock In press.

\bibitem{rubiola07uffc-am-to-pm-pollution}
E.~Rubiola and R.~Boudot.
\newblock The effect of {AM} noise on correlation phase noise measurements.
\newblock {\em IEEE Trans.\ Ultras.\ Ferroelec.\ and Freq.\ Contr.},
  54(5):926--932, May 2007.

\bibitem{rubiola08arxiv-amplifier-noise}
E.~Rubiola and R.~Boudot.
\newblock $1/f$ noise of {RF} and microwave amplifiers.
\newblock available soon on http://arxiv.org, 2008.

\bibitem{rubiola00rsi-correlation}
E.~Rubiola and V.~Giordano.
\newblock Correlation-based phase noise measurements.
\newblock {\em Rev.\ Sci.\ Instrum.}, 71(8):3085--3091, Aug. 2000.

\bibitem{rubiola02rsi-matrix}
E.~Rubiola and V.~Giordano.
\newblock Advanced interferometric phase and amplitude noise measurements.
\newblock {\em Rev.\ Sci.\ Instrum.}, 73(6):2445--2457, June 2002.
\newblock Also on the web site arxiv.org, document ar{X}iv:physics/0503015v1.

\bibitem{rubiola99rsi}
E.~Rubiola, V.~Giordano, and J.~Groslambert.
\newblock Very high frequency and microwave interferometric {PM} and {AM} noise
  measurements.
\newblock {\em Rev.\ Sci.\ Instrum.}, 70(1):220--225, Jan. 1999.

\bibitem{rubiola04rsi-amplifier}
E.~Rubiola and F.~Lardet-Vieudrin.
\newblock Low flicker-noise amplifier for 50\,{$\Omega$} sources.
\newblock {\em Rev.\ Sci.\ Instrum.}, 75(5):1323--1326, May 2004.
\newblock Free preprint available on the web site arxiv.org, document
  arXiv:physics/0503012v1, March 2005.

\bibitem{rubiola05josab-delay-line}
E.~Rubiola, E.~Salik, S.~Huang, and L.~Maleki.
\newblock Photonic delay technique for phase noise measurement of microwave
  oscillators.
\newblock {\em J. Opt. Soc. Am. B - Opt. Phys.}, 22(5):987--997, May 2005.

\bibitem{rubiola06mtt-photodiodes}
E.~Rubiola, E.~Salik, N.~Yu, and L.~Maleki.
\newblock Flicker noise in high-speed p-i-n photodiodes.
\newblock {\em IEEE Trans.\ Microw.\ Theory Tech.}, 54(2):816--820, Feb. 2006.
\newblock Preprint available on arxiv.org, document ar{X}iv:physics/0503022v1,
  March 2005.

\bibitem{rutman78pieee}
J.~Rutman.
\newblock Characterization of phase and frequency instabilities in precision
  frequency sources: Fifteen years of progress.
\newblock {\em Proc.\ IEEE}, 66(9):1048--1075, Sept. 1978.

\bibitem{salik04fcs-xhomodyne}
E.~Salik, N.~Yu, L.~Maleki, and E.~Rubiola.
\newblock Dual photonic-delay-line cross correlation method for the measurement
  of microwave oscillator phase noise.
\newblock In {\em Proc. Europ.\ Freq.\ Time Forum and Freq.\ Control Symp.
  Joint Meeting}, pages 303--306, Montreal, Canada, Aug.~23-27 2004.

\bibitem{salzenstein07appa-dual-delay-line}
P.~Salzenstein, J.~Cussey, X.~Jouvenceau, H.~Tavernier, L.~Larger, E.~Rubiola,
  and G.~Sauvage.
\newblock Realization of a phase noise measurement bench using cross
  correlation and double optical delay line.
\newblock {\em Acta Phys.\ Polonica A}, 112(5):1107--1111, Nov. 2007.

\bibitem{savchenkov04prl}
A.~A. Savchenkov, A.~B. Matsko, D.~Strekalov, M.~Mohageg, V.~S. Ilchenko, and
  L.~Maleki.
\newblock Low threshold optical oscillations in a whispering gallery mode caf2
  resonato.
\newblock {\em Phys.\ Rev.\ Lett.}, 93(24):1--4, Dec.~10, 2004.

\bibitem{shieh05ptl}
W.~Shieh and L.~Maleki.
\newblock Phase noise characterization by carrier suppression techniques in
  {RF} photonic systems.
\newblock {\em IEEE Photonic Technology Lett.}, 17(2):474--476, Feb. 2005.

\bibitem{maleki98ofc}
W.~Shieh, X.~S. Yao, L.~Maleki, and G.~Lutes.
\newblock Phase-noise chracterization of optoelectronic components by carrier
  suppression techniques.
\newblock In {\em Proc. Optical Fiber Comm. (OFC) Conf.}, pages 263--264, San
  Jos\'{e}, {CA}, May~23-25 1998.

\bibitem{vahala03nature}
K.~J. Vahala.
\newblock Optical microcavities.
\newblock {\em Nature}, 424:839--846, Dec.~10, 2003.

\bibitem{vanier:frequency-standards}
J.~Vanier and C.~Audoin.
\newblock {\em The Quantum Physics of Atomic Frequency Standards}.
\newblock Adam Hilger, Bristol, UK, 1989.

\bibitem{ieee99std1139}
J.~R. {Vig (chair.)}.
\newblock {\em {IEEE} Standard Definitions of Physical Quantities for
  Fundamental Frequency and Time Metrology--Random Instabilities ({IEEE}
  Standard 1139-1999)}.
\newblock {IEEE}, New York, 1999.

\bibitem{laser-tuning}
K.~Volyanskiy and L.~Larger.
\newblock Quadrature stabilization in the opto-electronic phase-noise
  measurement system by laser tuning.
\newblock Femto-st internal report, May 2008.

\bibitem{walls97uffc}
F.~L. Walls, E.~S. {Ferre-Pikal}, and S.~R. Jefferts.
\newblock Origin of $1/f$ {PM} and {AM} noise in bipolar junction transistor
  amplifiers.
\newblock {\em IEEE Trans.\ Ultras.\ Ferroelec.\ and Freq.\ Contr.},
  44(2):326--334, Mar. 1997.

\bibitem{yao00jlt}
X.~S. Yao, L.~Davis, and L.~Maleki.
\newblock Coupled optoelectronic oscillators for generating both {RF} signal
  and optical pulses.
\newblock {\em J. Ligtwave Technol.}, 18(1):73--78, Jan. 2000.

\bibitem{yao96josab}
X.~S. Yao and L.~Maleki.
\newblock Optoelectronic microwave oscillator.
\newblock {\em J. Opt. Soc. Am. B - Opt. Phys.}, 13(8):1725--1735, Aug. 1996.

\bibitem{delfyett03jqe}
T.~A. Yilmaz, C.~M. Depriest, A.~Braun, J.~Abeles, and P.~J. Delfyett.
\newblock Noise in fundamental and harmonic modelocked semiconductor lasers:
  Experiments and simulations.
\newblock {\em J. Quantum Electron.}, 39(7):838--849, July 2003.

\end{thebibliography}

\end{document}